\documentclass[ea]{agutex2015}
\usepackage[finalnew]{trackchanges}
\addeditor{KB}
\usepackage{amssymb}
\usepackage{url}
\usepackage{graphicx}
\setkeys{Gin}{draft=false}
\usepackage{lineno}
\authorrunninghead{BOWMAN ET AL.}

\titlerunninghead{BRAZILIAN CARBON BALANCE \copyright 2016 All Rights Reserved}

\authoraddr{Corresponding author: K.W. Bowman,
	Jet Propulsion Laboratory, California Institute of Technology
	Pasadena, CA, 91109, USA.
	(kevin.bowman@jpl.nasa.gov)}

\begin{document}

%\linenumbers

\title{Global and Brazilian carbon response to El Ni\~{n}o Modoki 2011-2010 }

%% ------------------------------------------------------------------------ %%
%
%  AUTHORS AND AFFILIATIONS
%
%% ------------------------------------------------------------------------ %%

%Use \author{\altaffilmark{}} and \altaffiltext{}

% \altaffilmark will produce footnote;
% matching \altaffiltext will appear at bottom of page.

\authors{K. W. Bowman,\altaffilmark{1,2} J. Liu \altaffilmark{1}, A. A. Bloom\altaffilmark{1}, N. C. Parazoo\altaffilmark{2},M. Lee\altaffilmark{1}, Z. Jiang\altaffilmark{5}, D. Menemenlis\altaffilmark{1}, M. M. Gierach\altaffilmark{1}, G. J. Collatz\altaffilmark{3}, K. R. Gurney\altaffilmark{4}, D. Wunch\altaffilmark{6}}

\altaffiltext{1}{Jet Propulsion Laboratory, California Institute of Technology, Pasadena, CA}
\altaffiltext{2}{Joint Institute for Regional Earth System Science and Engineer, University of California Los Angeles, CA}
\altaffiltext{3}{NASA Goddard Space Flight Center, Greenbelt, MD}
\altaffiltext{4}{Ecology Evolution and Environmental Science, Arizona State University, Tempe, AZ}
\altaffiltext{5}{National Center for Atmospheric Research, Boulder, CO}
\altaffiltext{6}{Department of Physics, University of Toronto, Toronto}

%% These dates will be inserted by the Publication Production Office during the typesetting process.

%\firstpage{1}

%\maketitle  %% Please note that for the copernicus2.cls this command needs to be inserted after \abstract{TEXT}

\keypoints{\item 2010 El Ni\~no led to significant droughts in Brazil  \item Biomass burning from Brazil dominated global biomass burning tendency (2011-2010) \item Brazilian net flux tendency -0.24 $\pm$ 0.11 PgC is mostly driven by biomass burning \item  Positive Brazilian GPP tendency observed from solar induced fluorescence measurements is balanced by respiration. }

\begin{abstract}
The \remove[KB]{central Pacific} El Ni\~{n}o Modoki in 2010 lead to historic droughts in Brazil.   We quantify the global and Brazilian carbon response to this event using the NASA Carbon Monitoring System Flux (CMS-Flux) framework.   Satellite observations of CO$_2$, CO, and solar induced fluorescence (SIF)  are ingested into a 4D-variational assimilation system driven by carbon cycle models to infer spatially resolved carbon fluxes including net ecosystem exchange, biomass burning, and gross primary productivity (GPP).  The  global net carbon flux tendency\add[KB]{, which is the flux difference 2011-2010 and is  positive for net fluxes into the atmosphere,} was estimated to be -1.60 PgC between 2011-2010 while the Brazilian tendency was  -0.24 $\pm$ 0.11 PgC\change[KB]{, which}{. This estimate} is broadly \add[KB]{within} the uncertainty of previous aircraft based estimates restricted to the Amazonian basin. The biomass burning tendency in Brazil was -0.24 $\pm$ 0.036 PgC, which implies a near-zero change of the net ecosystem production (NEP). The near-zero change of the NEP is the result of quantitatively comparable increase in GPP (0.34 $\pm$ 0.20) and respiration in Brazil. Comparisons of the component fluxes in Brazil to the global fluxes show a complex balance between regional contributions to individual carbon fluxes such as biomass burning, and their net contribution to the global carbon balance, e.g., the Brazilian biomass burning tendency is a significant contributor to the global biomass burning tendency but the Brazilian  net flux tendency is not a dominant contributor to the global tendency.  These results show the potential of multiple satellite observations to help quantify the spatially resolved response of productivity and respiration fluxes to climate variability. 
\end{abstract}

%% only used for copernicus2.cls
%\abstract{
% TEXT
% \keywords{TEXT}}

\begin{article}

\section{Introduction}  %% \introduction[modified heading if necessary]

In 2011, carbon dioxide (CO$_2$) measurements reached 391 ppm, which is about 40\% higher than preindustrial levels.  This level substantially exceeds the highest concentrations recorded in ice cores during the past 800,000 years \citep{Stocker:2013fk}.  These dramatic changes have led to a planetary radiative imbalance estimated at 0.58 $\pm$ 0.15 Wm$^{-2}$ from 2005-2010 at the top-of-the-atmosphere \citep{Hansen:2011nx}.  Changes in atmospheric temperature, hydrology, sea ice, and sea levels are attributed to climate forcing agents dominated by  CO$_2$ \citep{Santer:2013fk,Stocker:2013fk}.  While increases in atmospheric CO$_2$ are a result of fossil fuel emissions, about 55\% is removed through terrestrial and oceanic physical and biogeochemical processes \citep{Gloor:2010uq}.  The ``airborne fraction'' (AF) is the ratio of the observed  atmospheric CO$_2$ \remove[KB]{to the  CO$_2$} to the CO$_2$ emitted from anthropogenic emissions. While this trend in the airborne fraction has been remarkably stable, some studies have suggested the AF is changing while others dispute this conclusion \citep{JosepG.Canadell11202007,Knorr:2009fk,Le-Quere:2009fj,Gloor:2010uq}.  Nevertheless, the majority of carbon-climate models indicate that the AF  will likely change in the future as a result of carbon-climate feedbacks \citep{Jones:2013ys}.   Assessing these trends is challenging in part due to the significant interannual variability of the atmospheric growth rate linked to natural variability in the climate system \citep{Langenfelds:2002uq,Wang:2013kx}.  \cite{Wang:2013kx} showed that a 1$^{\circ}$C tropical temperature anomaly leads to a 3.5 $\pm$ 0.6 PgC yr$^{-1}$ CO$_2$ growth rate anomaly (1959-2011) ($r^2 \approx 0.5$).  \cite{Cox:2013fk} used this same relationship between tropical temperatures and CO$_2$ to apply an emergent linear constraint on the carbon-climate feedback factor--the so-called $\gamma$ parameter (GtC K$^{-1}$)--diagnosed from the C$^{4}$MIP Earth System Models (ESM).  

These studies, however, do not pinpoint the spatial drivers of the CO$_2$ atmospheric growth rate.  Resolving the source of atmospheric growth rate variability is an important step in assessing the underlying processes driving these changes.  Globally distributed atmospheric observations of CO$_2$ have played an important role in quantifying the spatial distribution of CO$_2$ fluxes, \citep[e.g.,][]{RAYNER:1996nx,Gurney:2002mz,Peters:2007rc,Chevallier:2010bs,Chevallier:2011fk,Ciais:2010cr,Peylin:2013fk}.  One of the fundamental challenges of these approaches to quantify CO$_2$ fluxes has been the sparsity of the surface observations, especially in South America and Africa \citep{PATRA:2003ve}.  Observations of CO$_2$ from meteorological and atmospheric composition sounders have good coverage over these areas.  However, they are primarily sensitive to free tropospheric CO$_2$, which is well-mixed, and consequently has been challenging to use \citep{Chevallier:2005fk,Engelen:2004nx,Engelen:2009ph,Nassar:2011ly}.  More recently, near-infrared (NIR) observations of column CO$_2$ from the Japanese GOSAT satellite have the potential to greatly improve our understanding regional carbon fluxes \citep{Chevallier:2007fv,Baker:2010cu,Hungershoefer:2010fk}.  GOSAT has been used to assess  CO$_2$ variability in  northern and southern latitudes \citep{Guerlet:2013cr,Wunch:2013fk,Parazoo:2013fk}, \add[KB]{and to} quantify CO$_2$ fluxes from megacities \citep{Kort:2012zr,Silva:2013fk} and biomass burning \citep{Ross:2013uq}.  More recently, GOSAT observations have been used to infer global surface CO$_2$ fluxes \citep[e.g.,][]{Basu:2013kx,Maksyutov:2013fk,Houweling:2015aa,Deng:2014fk,Deng:2016aa}.

We will focus on how the historic 2010  El Ni\~n{o} Modoki and the following La Ni\~n{a} impacted global fluxes generally and Brazilian fluxes in particular.  These years provide a useful--though limited--contrast in the changes in atmospheric CO$_2$ growth rate.  The atmospheric growth rate observed from the NOAA Mauna Loa site for 2010 was 2.36 ppm yr$^{-1}$ in contrast to 1.88 $\pm 0.11$ ppm yr$^{-1}$ for 2011.  \change[KB]{The detrended anomaly CO$_2$ growth rate}{The detrended  CO$_2$ growth rate anomaly}  relative to the 1959-2011 mean  was  0.7 and -0.3  PgC~yr$^{-1}$ in 2010 and 2011, respectively. 
% The tropical land-surface temperature anomaly was 0.36$^{\circ}$C and -0.12$^{\circ}$C for 2010 and 2011, respectively \citep{Wang:2013kx}.  
A critical regional climate pattern was a historic sea surface temperature (SST) anomaly in the tropical North Atlantic during 2010 that reached almost 1$^\circ$C, breaking the previous record in 2005.  \remove[KB]{This SST anomaly was partly a response to the historic El  Ni\~{n}o Modoki where the El Ni\~{n}o 4.0 index exceeded 1.0,  but was additionally coupled to a strong and persistent negative phase of the North Atlantic Oscillation (NAO)   \protect\citep{Hu:2011fk}} \add[KB]{This SST anomaly was partly a response to a strong central Pacific or El Ni\~{n}o Modoki, which  is associated with strong anomalous warming in the central tropical Pacific and cooling in the eastern and western tropical Pacific, where the  El Ni\~{n}o Modoki index  exceeded 1.0$^{\circ}$C \protect\citep{Ashok:2007fk,Ashok:2009eq}. This El Ni\~{n}o Modoki was the largest relative to the previous three decades \citep{Lee:2010uq}. \cite{Hu:2011fk} showed the anomalous warming was additionally coupled to a strong and persistent negative phase of the North Atlantic Oscillation (NAO).} As a consequence, the Intertropical Convergence Zone (ITCZ) shifted north resulting in historic droughts in the Amazon in 2010 \citep{Lewis:2011uq}.  

We will use the NASA Carbon Monitoring System Flux (CMS-Flux), which integrates satellite observations across the carbon cycle, to investigate the spatial drivers of these changes and attribute them to regional carbon cycle processes \citep{Liu:2014kx}.  Atmospheric observations include  xCO$_{2}$ from the GOSAT instrument, which will be described in Sec.~\ref{sec:gosat} and  CO  from MOPITT, which will be discussed in Sec.~\ref{sec:mopitt}.  These data along with satellite observations of the land-surface are ingested into the  CMS-Flux framework , which is described in Sec.~\ref{sec:cmsflux}.  The 4D-variational system used to infer global CO$_2$ and CO fluxes  is described in Sec.~\ref{sec:4dvar}.  The net CO$_2$ flux is the sum of gross fluxes including biomass burning, gross primary productivity, total ecosystem respiration, fossil fuel, etc.  We develop an attribution strategy discussed in Sec.~\ref{sec:attrib} that uses a combination of observations, e.g., solar-induced fluorescence measurements, to estimate individual fluxes and then infer unmeasured respiration fluxes as a residual term.  We quantify the global flux tendencies, i.e., the change in flux between 2011 and 2010, derived from CMS-Flux in Sec.~\ref{sec:results} and then apply this attribution strategy to assess the Brazilian carbon tendency in Sec.~\ref{sec:brazil}.  The limitations and potential of this approach for understanding the role of the El Ni\~{n}o Southern Oscillation (ENSO) on the interannual variability of CO$_2$ will be discussed in Sec.~\ref{sec:conc}.

\section{GOSAT satellite}
\label{sec:gosat}
The Japan Aerospace Exploration Agency (JAXA) Greenhouse gases Observing Satellite ``IBUKI'' (GOSAT) satellite was launched in January 2009.  \change[KB]{GOSAT orbits  a 666km altitude, 98$^{\circ}$ inclination, sun-synchronous orbit }{GOSAT is in a sun-synchronous orbit at a 666 km altitude and a 98$^{\circ}$ inclination} with a 12:49 p.m. nodal crossing time and a three-day (44-orbit) ground track repeat cycle \citep{Hamazaki:2005uq}.  \remove[KB]{The} GOSAT measures both along track ($\pm~20^{\circ}$) and cross-track ($\pm~35^{\circ}$) with an instantaneous field-of-view (IFOV) of 10.5 km.  The 5-point cross-track scan mode was used  between 4 April 2009 and 31 July 2010 providing observations separated by $\approx$ 158 km cross-track and $\approx$ 152  km along track. Afterwards, the mode was changed to 3-point leading to $\approx$ 263 km cross-track and $\approx$ 283 km along-track~\citep{M.Nakajima:2010kx,Crisp:2012uq}.  This satellite includes the  Thermal And Near-infrared Sensor for carbon Observation-Fourier Transform Spectrometer (TANSO-FTS), which measures spectrally resolved radiances in the 0.76, 1.6, 2.0, 5.6 and 14.3 $\mu$m \add[KB]{bands} \citep{Kuze:2009fk}. These radiances are used to infer profile and column measurements of  CO$_2$ and CH$_4$ as well as other atmospheric variables including tropospheric ozone~\citep{Parker:2011uq,Ohyama:2015ys,Oshchepkov:2013fk}.   There are nadir (over land) and glint (over ocean) modes as well ``high" and ``medium" gain settings to adjust for surface brightness conditions.  In order to mitigate potential systematic errors and biases between observational modes,  we only use nadir and ``high" gain observations.  

The inference of dry-column CO$_2$ (xCO$_2$) is calculated from the  NASA Atmospheric CO$_2$ Observations from Space (ACOS) retrieval algorithm \citep{Connor:2008yu,ODell:2012fk}.  Based upon an optimal estimation framework, this algorithm  adjusts the atmospheric state, which includes the vertical distribution of CO$_2$ and other physical parameters,  to minimizes the Euclidean $L^2$ norm difference between observed and calculated radiances subject to knowledge of the second-order statistics of that state.\add[KB]{The ACOS algorithm is currently the only approach that has been applied operationally to both GOSAT and OCO-2 data. As we anticipate extending our analysis into the OCO-2 period, we exclusively use those data products.} \remove[KB]{The use of ACOS products insures continuity between the xCO$_2$ from GOSAT and OCO-2.} We use xCO$_2$ processed with ACOS version 3.5b, which includes a consistently calibrated spectra (v150151) across the GOSAT period \citep{Osterman:2013kx}. Bias correction and data filtering follow the recommendations in the ACOS User's Guide. \add[KB]{In particular, observations contaminated with clouds and aerosols are removed.}  In addition to xCO$_2$, the ACOS algorithms provides a number of diagnostic parameters critical for constructing observation operators used in data assimilation, including the averaging kernel and a priori state vectors. \citep{Rodgers:imastp-00,Jones:2003qd}.  
From an inverse modeling standpoint, the ACOS retrieval is represented as an additive noise model
\begin{equation}
y_i = h_i(\mathbf{c}_i)+\epsilon_i
\label{eq:nm}
\end{equation}
where $y_i\in\mathbb{R}$ is the observed xCO$_2$ at location and time $i$,  $h_i:\mathbb{R}^{P}\rightarrow\mathbb{R}$ is the ACOS observation operator, ${\bf c}_{i}\in\mathbb{R}^{P}$ is the vertical profile of CO$_2$ observed by GOSAT, and $\epsilon_{i}$ is the observation error with variance
\begin{equation}
\sigma^{2}_i = E[\epsilon^2_i],
\label{eq:obserr}
\end{equation}
where $E[\cdot]$ is the expectation operator.The observation error explicitly incorporates random measurement error from GOSAT and implicitly includes model representation and transport error. The observation operator is 
\begin{equation}
h_i(\cdot) = x^{b}_{CO_2} + \mathbf{a}^{\top}_{i}(\cdot-\mathbf{c}^{b}_{i})
\label{eq:obsop}
\end{equation}
where $\mathbf{a}\in\mathbb{R}^{P}$ is the pressure-weighted averaging kernel, $\mathbf{c}^{b}_{i}\in\mathbb{R}^{P}$ is the a priori CO$_2$ concentration vertical profile, and $x^{b}_{CO_2}\in\mathbb{R}$ is the a priori xCO$_2$ defined by 
\begin{equation}
x^{b}_{CO_2} = \frac{\int_{0}^{\mathrm{TOA}} c^b(z) N_{d}(z) \mathrm{d}z}{\int_{0}^{\mathrm{TOA}} N_{d}(z) \mathrm{d}z}
\label{eq:den}
\end{equation}
where $c^b$ is the continuous form of the a priori profile as a function of altitude $z$, $N_d$ is the number density, and $\mathrm{TOA}$ is the top-of-the-atmosphere.  

In addition to xCO$_2$, GOSAT measures solar-induced fluorescence (SIF) exploiting the Fraunhofer lines outside the oxygen $O_2$-A band in the 756--759 nm and 770.5--774.5 nm range~\citep{Frankenberg:2011mz}. These will be used to estimate Gross Primary Productivity (GPP) in Sec.~\ref{sec:gpp}.

\begin{figure}
	\centering
	\includegraphics[width=15cm]{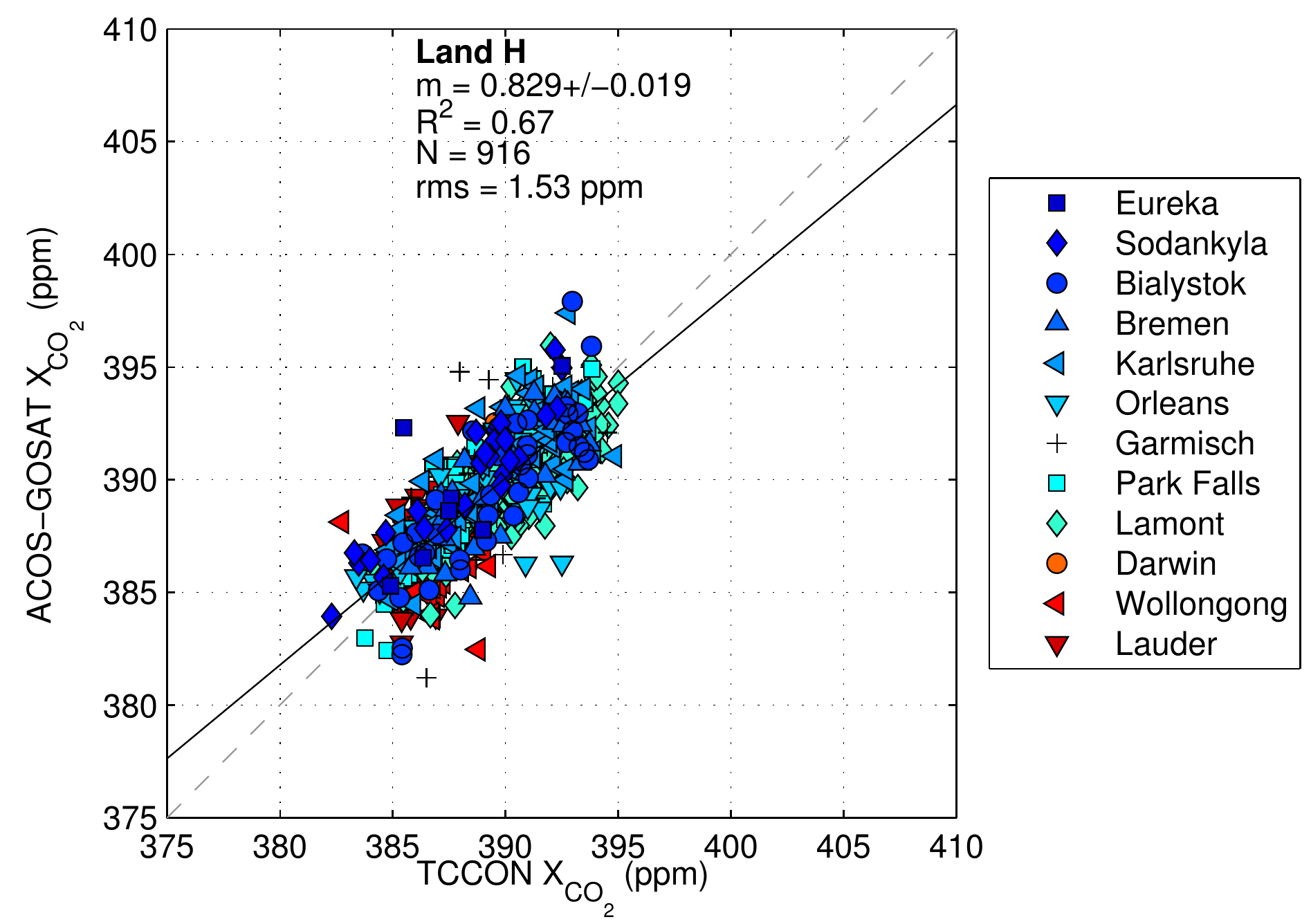}
	\caption[tccon-gosat]{Regression of GOSAT ACOS v3.5b xCO$_2$ retrievals against TCCON data for (a) 2010--2011. Only high gain (H) land data are used.  The slope is denoted by $m$ and the number of points by $N$.  The median difference between GOSAT and TCCON is 0.32 ppm.  }
	\label{fig:tccon-gosat}
\end{figure}

\add[KB]{The primary independent set of measurements to test  GOSAT retrievals is the Total Carbon Column Observing Network (TCCON) \protect{\citep{Wunch:2011uq,Deutscher:2014tp,Griffith:2014oh,Griffith:2014ph,Hase:2014qh,Kawakami:2014yg,Kivi:2014mk,Morino:2014fd,Notholt:2014rc,Sherlock:2014mb,Strong:2014rp,Sussmann:2014gn,Warneke:2014vs,Wennberg:2014mo,Wennberg:2014vp}} which is a suite of ground-based high resolution uplooking Fourier Transform Spectrometers.  Previous studies have reported that for several different retrieval algorithms, the global bias between GOSAT  and TCCON xCO$_2$ column retrievals is generally within 1 ppm from 2009-2011 \citep{Oshchepkov:2013fk}. These algorithms, however, are frequently updated, (e.g., ACOS 2.9 was used in the comparison whereas this study uses ACOS v3.5b),  and therefore the comparisons are best treated as a snapshot.  The covariation between collocated  ACOS v3.5b and TCCON data for 2010--2011 are shown in Fig.~\ref{fig:tccon-gosat}. The coincidence criteria was  $\pm$2.5$^{\circ} $ latitude and 
$\pm$5.0$^{\circ}$ longitude in the Northern Hemisphere. For the Southern Hemisphere poleward of 25$^{\circ}$S,  the coincidence criteria of $\pm$10$^{\circ}$  latitude and $\pm$60$^{\circ}$ longitude was used. The root-mean-square (RMS) is 1.53 ppm, the R$^2$ is 0.67 and the median bias is 0.32 ppm. These quantities  do not change significantly between 2010 and 2011.  The slope is  approximately 0.83 and indicates that GOSAT will likely underestimate relative to TCCON the  seasonal cycle amplitude in the Northern Hemisphere, which is where most of the sites are located \citep{Lindqvist:2015kq}.  These comparisons for the ACOS v3.5b algorithm are quantitatively consistent with  \cite{Kulawik:2016fj}, which further benefited from TCCON stations in the Ascension Islands and Manaus, Brazil that started operations after 2011.  \cite{Frankenberg:2016fk} showed for ocean  glint retrievals, which use the same ACOS algorithm, have a bias of -0.06 ppm and R$^2$=0.85 relative to High-Performance Instrumented Airborne Platform for Environmental Research (HIAPER) Pole-to-Pole Observations (HIPPO) flights from January 2009 through September 2011.    Indirect comparisons of transport model concentration data to aircraft measurements can provide additional insight as discussed in Sec.~\ref{sec:brazil}.  } 

\section{MOPITT}
\label{sec:mopitt}
The MOPITT instrument \citep{Drummond:2010fk} was launched on 18 December 1999 on NASA's Terra spacecraft in a sun-synchronous polar orbit at an altitude of 705 km with an equator crossing time of 10:30 a.m. local time. With a footprint of 22 km $\times$ 22 km, the instrument has a 612 km cross-track scan and achieves near global coverage every 3 days. Carbon monoxide (CO) is inferred from a  gas correlation radiometer that measures thermal emission in the 4.7 $\mu$m  and 2.3 $\mu$m regions \citep{Deeter:2003ve}.  

The combination of thermal IR and NIR channels provides MOPITT the unique capability to infer near-surface CO \citep{Worden:2010bh} and its long time series has been used to infer decadal trends \citep{Worden:2013fk}. We use MOPITT V6 profiles, which have better geolocation and meteorological inputs than previous versions \url{https://www2.acom.ucar.edu/mopitt/publications}.   These CO observations will be used to infer CO$_2$ biomass burning emissions in Sec.~\ref{sec:bb}. 
%TODO Need to verify what version of MOPITT we are actually using. 
\section{CMS-Flux framework}
\label{sec:cmsflux}
\begin{figure}
    \centering
	\includegraphics[width=15cm]{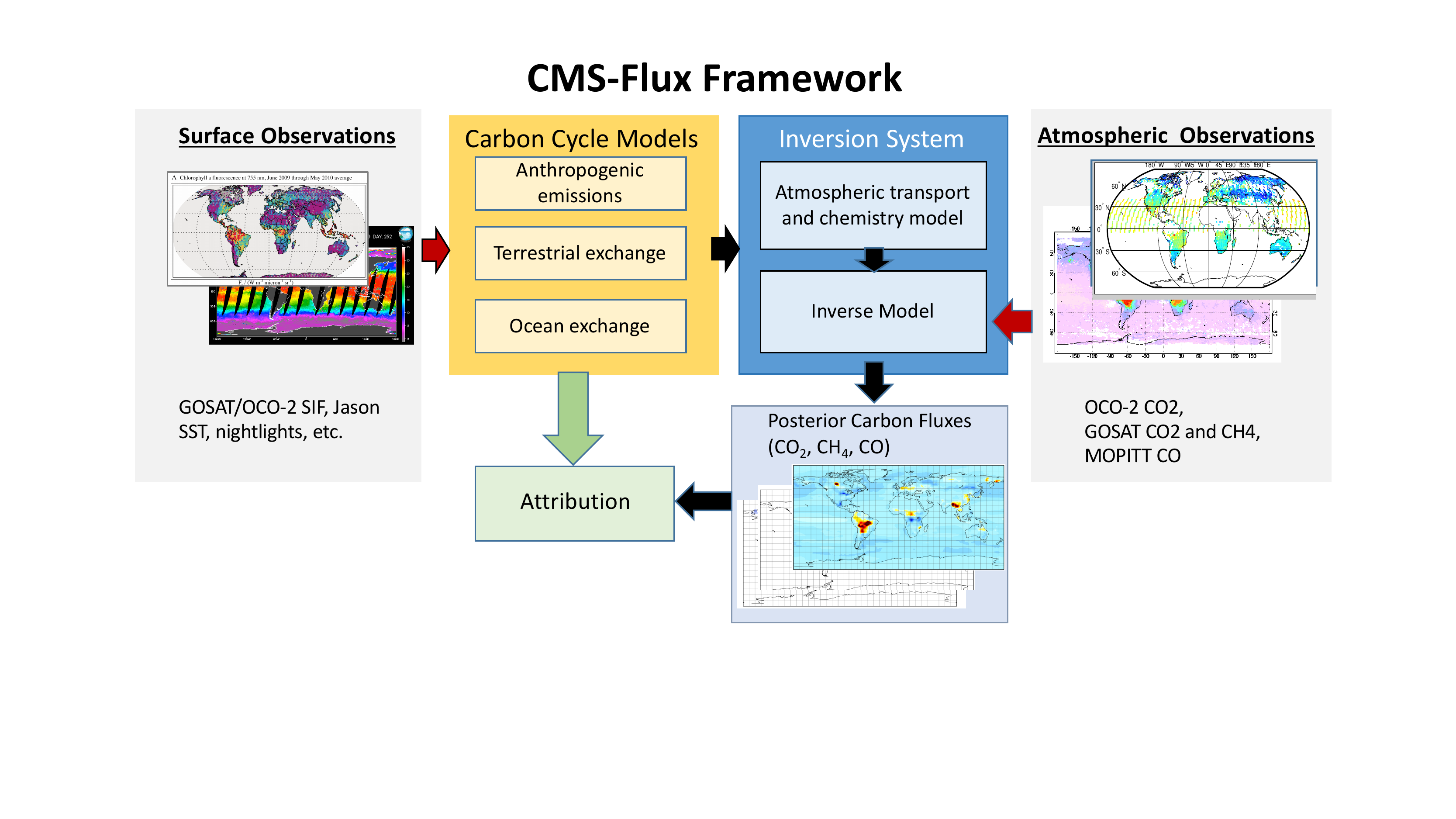}
	\caption[cms-framework]{Carbon Monitoring System Flux (CMS-Flux) Framework. Satellite observations of surface data are integrated into a suite of anthropogenic (FFDAS), ocean (ECCO2-Darwin), and terrestrial (CASA-GFED) carbon cycle models. These are in turn used to compute surface fluxes that drive a chemistry and transport model (GEOS-Chem). Atmospheric observations of CO$_2$, CO, CH$_4$ are ingested into an inverse model that computes  posterior estimates of carbon surface fluxes.  The combination of fluxes is used to attribute carbon and then reconcile those differences with prior carbon cycle models. }
	\label{fig:cms-framework}
\end{figure}

The CMS-Flux framework is shown in Fig.~\ref{fig:cms-framework}.  \remove[KB]{Satellite observations of surface data are integrated into a suite of carbon cycle models for anthropogenic \protect{(Sec.~\ref{sec:ffdas})}, ocean (Sec.~\ref{sec:ecco2}), and terrestrial (Sec.~\ref{sec:te}).} \add[KB]{Satellite observations of surface data are integrated into a suite of  anthropogenic \protect{(Sec.~\ref{sec:ffdas})}, ocean (Sec.~\ref{sec:ecco2}), and terrestrial (Sec.~\ref{sec:te}) carbon cycle models.}
These are in turn used to compute surface fluxes that drive a chemistry and transport model (GEOS-Chem) (Sec.~\ref{sec:geos-chem}). Atmospheric observations of CO$_2$ (Sec.~\ref{sec:gosat}), CO (Sec.~\ref{sec:mopitt}), CH$_4$ (not shown) are ingested into a 4D-variational inverse model (Sec.~\ref{sec:4dvar}) that computes a posterior estimate of carbon surface fluxes along with uncertainties \add[KB]{where the anthropogenic and oceanic fluxes are treated as fixed priors}. We then evaluate the accuracy of posterior fluxes relative to the prior by comparing the CO2 concentrations forced by either posterior or prior fluxes against independent data (Sec.~\ref{sec:brazil}). The use of multiple species facilitates the attribution of net carbon fluxes to component terms (Sec.~\ref{sec:attrib}).   The architecture is fairly flexible allowing for integration of other carbon cycle models, \cite[e.g.,][]{Schwalm:2015fk} as well as inversion methodologies~\citep{Liu:2016ab}.

\subsection{Fossil Fuel Data Assimilation System}
\label{sec:ffdas}
Fossil fuel emissions estimates are based upon data  at national and annual scales, and  disaggregated by proxy data. The Fossil Fuel Data Assimilation System (FFDAS), which ingests  Defense Meteorological Satellite Program (DMSP) night lights, national fossil fuel, a new power plant database (Ventus), and population \citep{Rayner:2010ys},  produced emissions and uncertainties on a $0.1^\circ\times 0.1^\circ$ grid for the 1997--2011 period \citep{Asefi-Najafabady:2014aa}.

\subsection{ECCO2-Darwin ocean biogeochemical assimilation system}
\label{sec:ecco2}
The ECCO2-Darwin estimates are based on 
(i) the Estimating the Circulation and Climate of the Ocean, Phase II (ECCO2) project, which provides global, eddying, data-constrained
estimates of physical ocean variables \citep{men05a,men08} and (ii) the Darwin
project, which provides ocean ecosystem variables \citep{fol07,fol11}. The 4D-variational approach insures 
inherent conservation properties of ECCO2 estimates over the assimilation window and are therefore particularly well
suited for application to ocean carbon cycle studies
\citep[e.g.,][]{fle06,fle07,gru09,man09,man11,man13}. Together, ECCO2 and
Darwin provide a time-evolving physical and biological environment for carbon
biogeochemistry \citep{dut11}. 

As part of CMS Phase II, air-sea gas exchange coefficients and initial
conditions of dissolved inorganic carbon, alkalinity, and oxygen were adjusted
using the Green's function approach of \cite{men05} in order to optimize
modeled air-sea CO$_2$\ fluxes.  Data constraints included observations of
carbon dioxide partial pressure (pCO$_2$) for 2009--2010, global air-sea
CO$_2$\ flux estimates, and the seasonal cycle of the Takahashi atlas \citep{Takahashi:2009fk}.  Green's Functions
include simulations that
start from different initial conditions as well as experiments that perturb
air-sea gas exchange parameters and the ratio of particulate inorganic to
organic carbon.   The Green's Functions approach yields a linear combination of
these sensitivity experiments that minimizes model-data differences.  The
resulting initial conditions and gas exchange coefficients are then used to
integrate the ECCO2-Darwin model forward.  
Despite only  six control parameters, the adjusted simulation is significantly closer to independent observations.   For example, the root-mean-square difference with observed
alkalinity decreased from 48.8~$\mu$mol/kg for the baseline simulation to
28.9~$\mu$mol/kg for the adjusted simulation, even though alkalinity
observations were not used as data constraints.  
In addition to reducing
biases relative to observations, the adjusted simulation exhibits smaller
model drift than the baseline.  For example, the volume-weighted drift
reduction in the top 300~m is 12.5\% for nitrate and 30\% for oxygen.  
This work, described in detail in \cite{Brix:2015fk}, resulted in ECCO2-Darwin Version-2
air-sea carbon flux estimates.
%A further improvement of the ECCO2-Darwin air-sea carbon flux estimates
%resulted by adopting the air-sea CO$_2$ exchange parameterization of
%\citep{kra06}, which varies linearly as opposed to quadratically with wind
%speed.  The resulting air sea flux estimates, called Version-3, further reduce
%model-data differences, and form the basis of the ocean work proposed herein.
\subsection{Terrestrial Ecosystem}
\label{sec:te}
The Carnegie-Ames-Stanford-Approach-Global Fire Emissions Database version 3 (CASA-GFED3)  \citep{Randerson:1996fk,Werf:2006qf,Werf:2010bh} ingests data including air temperature, precipitation, incident solar radiation, a soil classification map, and a number of satellite derived products including burned area, fractional woody cover, and fraction of photosynthetically active radiation absorbed (fAPAR). For these simulations meteorological data (temperature, precipitation, solar radiation) are taken from the GMAO Modern-Era Retrospective Analysis for Research and Applications (MERRA) \citep{Rienecker:2011qf}. 
CASA-GFED3 is run at monthly time steps with 0.5$^{\circ}$ spatial resolution. fAPAR is derived from Normalized Difference Vegetation Index (NDVI) \citep{Tucker:2005vn} according to the procedure of \cite{Los:2000fk}. 
The model output includes net primary production (NPP), heterotrophic respiration (RH), and fire emissions. Wildfire emissions were disaggregated from monthly to quasi-daily using the eight-day Moderate Resolution Imaging Spectrometer (MODIS)  MYD14A2 Active Fire Product (http://modis-fire.umd.edu/).
Using 3 hourly MERRA air temperature and incident solar radiation, monthly fluxes were disaggregated into 3 hourly gross biological fluxes and added to produce the 3 hourly net carbon flux to the atmosphere according to the approach of \cite{Olsen:2004yu}.  MERRA-driven CASA-GFED3 carbon fluxes have been used in a number of atmospheric CO$_2$ transport studies \citep[e.g.,][]{Campbell:2008ve,KAWA:2010kx,Hammerling:2012uq}. 

In order to have a consistent global carbon balance,  the annual global NEE from CASA-GFED is scaled to match the residual carbon from the sum of the  atmospheric growth rate, ocean, and fossil fuel estimates following the approach from the Global Carbon Project (GCP) \citep{Le-Quere:2009fj}.

\subsubsection{Parametric uncertainty}
To estimate uncertainties in CASA-GFED3 carbon fluxes we first performed sensitivity simulations in which model parameters were incrementally changed. These studies revealed that fluxes were highly sensitive to the first order parameters and driver data that determine NPP  (potential light use efficiency, fAPAR, incident solar radiation, temperature and moisture limitation scalars). Net fluxes were also highly sensitive to parameters associated with the temperature and moisture responses of heterotrophic respiration. Sensitivity of respiration to  carbon turnover rate parameters was low. Based on literature review the impacts of uncertainties in parameters and driver data of GPP were combined and estimated as a standard deviation of 20\% in the mean GPP (102-152 PgC/yr ) consistent with the range of values reported from atmospheric inversion optimizations, model intercomparisons, and empirically-driven data-based models \citep[e.g.,][]{Cramer:1999zr,Kaminski:2002ys,Jung:2011kx}.  Uncertainty in the temperature response of respiration was represented as a standard deviation of 20\% variability in the $Q_{10}$  (e.g., the  respiration sensitivity factor for a 10$^{\circ}$C  increase)  for temperature ranges (1.2-1.8) consistent with other published estimates \citep[e.g.,][]{Kaminski:2002ys,Mahecha:2010zr}.  The moisture response of respiration is not well defined in the literature and was prescribed at a standard deviation of 20\% in the moisture scalar (proportional with respiration) as a plausible value.  The CASA-GFED3 model was initially spun up for 1000 years to equilibrium with constant parameters and mean seasonal cycles of meteorology and satellite based fAPAR. Using a Monte Carlo approach sets of parameters were selected for each ensemble run and spun up for 200 years to the start of the variable data time series (1997-2011) and flux uncertainties around the mean were estimated. We also included uncertainties in fractional woody cover and woody mortality but while these uncertainties had a large influence on biomass estimates, they had virtually no effect on the fluxes.

\subsection{GEOS-Chem}
\label{sec:geos-chem}

GEOS-Chem (\url{http://www.geos-chem.org/}) is a global chemical transport model (CTM) driven by meteorological fields from the NASA Goddard Earth Observing System (GEOS) data assimilation system of the  Global Modeling and Assimilation Office (GMAO) \citep{Rienecker:2011qf}. We use GEOS Version 5 (GEOS-5) meteorology aggregated to $4^{\circ} \times 5^{\circ}$\citep{Bey:2001fk}, which is archived with a temporal resolution of 6 hours except for surface quantities and mixing depths that have temporal resolution of 3 hours.  Convective transport in GEOS-Chem is computed from the convective mass fluxes in the meteorological archive, as described by \cite{Wu:2007vn}.

 The simulation of CO$_2$ was originally implemented by \cite{Suntharalingam:2004cr}.  \cite{Nassar:2010zr} incorporated a number of updates including spatially explicit emissions from shipping,  aviation, and a chemical source, which is based on the oxidation of carbon monoxide (CO), methane (CH$_4$) and non-methane volatile organic carbons (NMVOCs) throughout the troposphere.  \cite{Nassar:2011ly} used this version to estimate coarse-resolution surface fluxes  constrained by mid-tropospheric CO$_2$ from the Tropospheric Emission Spectrometer \citep{Bowman:2006fk,Kulawik:2010kx}. The shipping, aviation, and chemical source terms are incorporated into the CMS-Flux carbon cycle described previously.  
 
 The GEOS-Chem adjoint (\url{http://wiki.seas.harvard.edu/geos-chem/index.php/GEOS-Chem_Adjoint}) described originally in \cite{Henze:2007fk} has been applied to a variety of fields including the estimation of inorganic fine particles (PM2.5) precursor emissions over the United States \citep{Henze:2009dk},  CO emissions \citep{Kopacz:2009db,Kopacz:2010et}, ozone assimilation \citep{Singh:2011kx}, and attribution of direct ozone radiative forcing \citep{Bowman:2012fk}. The development and application of the GEOS-Chem CO$_2$ adjoint is described in \cite{Liu:2014kx} and \cite{Deng:2014fk}.  

\subsection{Variational framework}
\label{sec:4dvar}
The Bayesian approach for inferring spatially-resolved fluxes informed by prior knowledge of the state is implemented in a 4D-variational framework.  Variational systems have been widely used to estimate sources of  CO$_2$ \citep[e.g.,][]{Kaminski:2002ys,Chevallier:2005fk,Rayner:2005tg,Ciais:2010cr,Baker:2010cu,Basu:2013kx,Deng:2014fk}. \remove[KB]{Adjoint-based variational approach differs from so-called ``analytic Bayesian" methods only in that variational systems can typically compute fluxes at orders-of-magnitude higher resolution  \protect\citep[e.g.,][]{RAYNER:1996nx,Gurney:2002mz,Jones:2003qd,Kopacz:2009db,Nassar:2011ly} but with the concomitant challenge of calculating posterior uncertainties. } \add[KB]{Adjoint-based variational approaches differs from so-called ``analytic Bayesian" methods  in that variational systems  typically compute fluxes at orders-of-magnitude higher resolution  \protect\citep[e.g.,][]{RAYNER:1996nx,Gurney:2002mz,Jones:2003qd,Kopacz:2009db,Nassar:2011ly}. Variational methods generally use iterative numerical techniques, e.g., conjugate-gradient, to compute an estimate and its uncertainty\citep{Bousserez:2015fk}. }

Bayes' Theorem provides a probabilistic framework to reduce uncertainty in the surface fluxes given measurements related to those fluxes through the following \citep{Papoulis:prvsp-84}:
\begin{equation}
p(\mathbf{x}|\mathbf{y}) = \frac{p(\mathbf{y}|\mathbf{x})p(\mathbf{x})}{p(\mathbf{y})}.
\label{eq:bayes}
\end{equation}
Under Gaussian assumptions, the a priori or ``background" distribution  $p(\mathbf{x})$ can be described uniquely by the 
 a priori vector and covariance matrix, ${\bf{x}}_{b}\in\mathbb{R}^{N}$ and $\mathbf{B}\in\mathbb{R}^{N\times N}$  ($N$ is the number of fluxes) respectively:
\begin{eqnarray}
{\bf{x}}_{b} & = & E[\bf{x}] \\
{\bf{B}} & = &E[\bf{x}\bf{x}^{\top}].
\label{eq:B}
\end{eqnarray}
Under Gaussian assumptions, $p({\bf y}|{\bf x})$ can be \change[KB]{related}{derived} from the observation error in Eq.~\ref{eq:obserr}.  The observational uncertainty can be combined with  a prior knowledge of the surface fluxes through Eq.~\ref{eq:bayes}
to define a cost function \citep{Lewis:2006uq,Jazwinski:2007fk}:
\begin{equation}
\label{eq:cf}
{\cal J}({\bf{x}}) = \frac{1}{2}||{\bf{y}} - {\cal H}({\bf{x}})||_{{{\bf{R}}^{ - 1}}}^2 + \frac{1}{2}\left\| {{\bf{x}} - {{\bf{x}}_b}} \right\|_{{{\bf{B}}^{-1}}}^2
\end{equation}
where ${\bf{y}}$ is a vector of GOSAT observations whose elements, $[{\bf{y}}]_{i} = y_{i}$, are defined in Eq.~\ref{eq:nm}, $\bf{R}\in\mathbb{R}^{M\times M}$ is the observational error covariance matrix with diagonal elements \remove[KB]{are} $[{\bf{R}}]_{i}  = \sigma^{2}_{i}$ defined in Eq.~\ref{eq:obserr}.The operator ${\cal H}:\mathbb{R}^{N}\rightarrow\mathbb{R}^{M}$ relates surface fluxes to each observation through the composite:
\begin{equation}
[{\cal H}]_{i} = (h_{i}\cdot {\cal M}_{i})({\bf x})
\end{equation} 
where $h_{i}$ is the observation operator defined in Eq.~\ref{eq:obsop} for the $i^{\mathrm{th}}$ observation, and ${\cal M}:\mathbb{R}^{N}\rightarrow\mathbb{R^{P}}$ is the GEOS-Chem transport operator 
\begin{equation}
{\bf c}_{i} = \mathcal{M}_{t_{0}\rightarrow t_{i}}({\bf x},{\bf c}_{0}),
\end{equation} 
\add[KB]{which} relates the surface fluxes to the vertical CO$_2$ profile viewed by GOSAT (Eq.\ref{eq:obsop}).  The initial conditions of the operator are defined at GEOS-Chem CO$_2$ concentrations, ${\bf c}_{0}$ at time $t_{0}$.
The maximum a posterior (MAP) estimate of the surface fluxes is 
\begin{equation}
\label{eq:mcf}
\hat{\bf x} = \min_{{\bf x}} {\cal J}({\bf x}),
\end{equation}
\add[KB]{which} represents the optimal balance of a prior knowledge of surface fluxes and the new information provided by the data.  The solution to Eq.~\ref{eq:mcf} requires the derivative of the cost function (Eq.~\ref{eq:cf})
\begin{equation}
\label{eq:grad}
\lambda  = {\nabla _{\bf{x}}}{\cal J}({\bf{x}}) = {\left( {\frac{{\partial {\cal M}}}{{\partial {\bf{x}}}}} \right)^ \top }{{\bf{h}}^ \top }{{\bf{R}}^{ - 1}}\left( {{\bf{y}} - {\cal H}({\bf{x}})} \right) + {{\bf{B}}^{ - 1}}\left( {{\bf{x}} - {{\bf{x}}_b}} \right)
\end{equation}
where $[\mathbf{h}]_i=h_i$, and \({\left( {{{\partial {\cal M}} \mathord{\left/
 {\vphantom {{\partial {\cal M}} {\partial {\bf{x}}}}} \right.
 \kern-\nulldelimiterspace} {\partial {\bf{x}}}}} \right)^ \top }\) and $\mathbf{h}^{\top}$ are adjoint operators \citep{Errico:1997kx}. For this study, the gradient in Eq.~\ref{eq:grad} is ingested into the Limited-memory Broyden-Fletcher-Goldfarb-Shannon (L-BFGS) quasi-Newton minimization technique to solve Eq.~\ref{eq:mcf} \citep{Byrd:1994fk,Zhu:1997nr}. \add[KB]{A necessary condition for convergence in \protect Eq.~\ref{eq:mcf}  is $\lambda\rightarrow 0$ where zero is computed to double precision.}  

The uncertainty of the MAP estimate or posterior uncertainty in Eq.~\ref{eq:mcf} is the covariance of the innovation, $\tilde{{\bf x}}= {\bf x}_{a}-\hat{\bf x}$, \citep{Lewis:2006uq} 
\begin{equation}
\label{eq:post}
{\widehat {\bf{P}}_x} = E[{\bf{\tilde x}}{{\bf{\tilde x}}^ \top }] = \left({\nabla ^2}{\cal J}({\bf{x}})\right)^{-1} = {\left( {{{\cal H}^ \top }{{\bf{R}}^{ - 1}}{\cal H} + {{\bf{B}}^{ - 1}}} \right)^{ - 1}}.
\end{equation}
For large order systems, Eq.~\ref{eq:post} can not be computed explicitly.  Both analytic and stochastic approaches have been used to estimate the posterior covariance, \citep[e.g.,][]{Nocedal:2006uq,Chevallier:2007fv}.  We use a Monte Carlo based method, which is described in \cite{Liu:2014kx} though hybrid techniques that incorporate both stochastic and analytic approaches have been investigated \citep{Bousserez:2015fk}.
\section{Attribution strategy}
\label{sec:attrib}
\subsection{Flux decomposition}
The net carbon flux at any given grid box can be expressed as the sum of component fluxes:
\begin{equation}
{F^\uparrow}(x,y,t) = ({F_{ff}} + {F_o} + {F_{bb}} - {F_{NEP}} + {F_{chem}})(x,y,t)
\label{eq:flux}
\end{equation}
where $F_{ff}$ is the fossil fuel flux, $F_o$ is the ocean flux, $F_{bb}$ is the biomass burning flux, $F_{NEP}$ is the net ecosystem production (NEP), $F_{chem}$, which is discussed in Sec.~\ref{sec:geos-chem}, is the chemical source in the overhead column as a function of location $(x,y)$ and $t$ . The sign convention assumes positive fluxes into the atmosphere. Consequently, positive NEP fluxes are negative in this convention. Similarly, the change in flux at any grid point can be expressed as:
\begin{equation}
{\delta F^\uparrow}(x,y,t) = ({\delta F_{ff}} + {\delta F_o} + {\delta F_{bb}} - {\delta F_{NEP}} + {\delta F_{chem}})(x,y,t)  
\label{eq:dflux}
\end{equation}
The control vector, $\mathbf{x}$,  estimated from the maximum a posterior solution of Eq.~\ref{eq:bayes} is modeled as NEP fluxes. The total posterior flux estimate is treated as the sum of the posterior NEP with the other component fluxes.  
The NEP flux can be further decomposed into 
\begin{equation}
  F_{NEP} = F_{GPP}-F_{R}
  \label{eq:NEP}
\end{equation}
where $F_{GPP}$ is the gross primary productivity and $F_{R}$ is the sum of heterotrophic and autotrophic respiration.  

Given independent estimates and uncertainties  of  component fluxes, unmeasured fluxes can be inferred as a residual. The uncertainty in the residual flux is based on the propagation of uncertainty from the independent measurements.  The attribution strategy is to estimate the net carbon flux using xCO$_2$ and the biomass burning derived from MOPITT CO emissions.  The fossil fuel and ocean estimates from FFDAS and ECCO2-Darwin along with the chemical source will be used to infer the NEP fluxes.  The GPP was derived from solar induced fluorescence (SIF) (Sec.~\ref{sec:gpp}). From the NEP and GPP, the total respiration, $F_{R}$, can be inferred.  For terrestrial flux tendencies in tropical regions, the changes in fossil fuel, ocean, and chemical source derived from FFDAS, ECCO2-Darwin, and GEOS-Chem, respectively,   are small relative to the NEP and BB flux tendencies.   Nevertheless, the residual flux may contain other sources not related to the total respiration and consequently, the attribution of the residual flux to the respiration flux has an additional level of  uncertainty.  
%{\color{blue} 
%\begin{itemize}
%\item Decomposition of flux into constituent components
%\item Decomposition of flux change into constituent components
%\item Inference of NEP and Rh from estimated components.  
%\item Propagation of uncertainty
%\end{itemize}
%}
\subsection{Biomass burning}
\label{sec:bb}
%\begin{itemize}
%\item Discuss the inversion approach for MOPITT CO
%\item cite references to Zhe and variational approach above
%\item Description of how biomass burning is converted to combustion
%\item cite van der werf
%\item Description of uncertainty calculation, Monte Carlo approach {\color{blue} Bloom}
%\end{itemize}

The estimate of CO emissions follows the variational framework in Sec.~\ref{sec:4dvar} and has been extensively documented \citep{Kopacz:2009db,Kopacz:2010et,Jiang:2015aa} \change[KB]{and}{including} its sensitivity to model error \citep{Jiang:2011kx,Jiang:2013uq}. Following \cite{Jiang:2011kx}, each month is estimated independently with initial conditions supplied by a sub-optimal Kalman filter \citep{Parrington:2008nr}. The configuration for the CO inversion follows \cite{Jiang:2013uq} where the control vector for CO emissions combines the combustion CO sources (fossil fuel, biofuel and biomass burning) with the CO from the oxidation of biogenic NMVOCs for each grid box and the source of CO from the oxidation of methane is estimated separately as an aggregated global source, assuming an a priori uncertainty of 25\%. 

Most of the CO emitted in tropical regions is driven by biomass burning \citep{Werf:2010bh}. The spatial attribution of biomass burning is based on burned area described in \cite{Giglio:2013fk}.  The contribution of combustion to carbon emissions \citep[e.g., ][]{Werf:2008ud,Worden:2012fk,Silva:2013fk,Bloom:2015kx} has been inferred from estimated ratios of CO to CO$_2$ and CH$_4$ as well as other gases \citep{Akagi:2011ve,Andreae:2001fk,Leeuwen:2014ve}. The estimate of CO$_2$ from CO follows the methodology in \citep{Bloom:2015kx}.  
The uncertainty in the CO$_2$ flux is the product of the uncertainty in the CO$_2$-to-CO ratio  and  the inferred CO emissions. The uncertainty in the CO emissions from the 4D-var assimilation follows the Monte Carlo approach in \cite{Liu:2014kx}.  For each gridbox and at each time-step, the  CO$_2$:CO ratio is 
\begin{equation}
r_{CO2:CO}=\frac{\sum^{6}_{i=1}e^{i}_{CO2}}{\sum^{6}_{i=1}e^{i}_{CO}}
\label{eq:ratio}
\end{equation}
where $e^{i}_{CO2}$and $e^{i}_{CO}$ are the GFED version 4 bottom-up estimates of CO$_2$ and CO emissions from six biomass burning sectors~\citep{Werf:2010bh}.  
The uncertainty in $r_{CO2:CO}$ at each grid box is  calculated with a Monte Carlo approach where the $j^{\mathrm{th}}$ sample is computed as follows: 
\begin{equation}
  r^{j}_{CO_2:CO} = \frac{\sum_{i=1}^{6} \left( f^{i,j}_{CO_2} e^{i}_{CO_2} - (1-f^{i,j}_{CO}) e^{i}_{CO})\right)}{ f^{i,j}_{CO} e^{i}_{CO}}
\end{equation}
where $e^{i}_{CO_2}$ and $e^{i}_{CO}$ are the total CO$_2$ and CO emissions for the i\textsuperscript{th} sector (see Table~\ref{tab:emfac}), and $f^{i,j}_{CO_2}$ and $f^{i,j}_{CO}$ are samples from the respective uncertainty factors. The addition of  $(1-f^{i,j}_{CO})e^{i}_{CO}$ in the numerator expresses the first-order anti-correlation between observed CO and CO$_2$ emission factors \citep[e.g.,][]{Yokelson:2007fk,Andreae:2001fk,Smith:2014fk}.
The samples from the uncertainty factors are computed as follows: 
\begin{eqnarray}
	f^{i}_{CO_2} &=& \left(\frac{\frac{\epsilon^{i}_{CO_2} + \sigma^{i}_{CO_2}}{\epsilon^{i}_{CO_2}}+\frac{\epsilon^{i}_{CO_2}}{\epsilon^{i}_{CO_2}-\sigma^{i}_{CO_2}}}{2}\right)^{N_{a}(0,1)} \label{eq:co2coeff}\\
	f^{i}_{CO} &=& \left(\frac{\frac{\epsilon^{i}_{CO}+\sigma^{i}_{CO}}{\epsilon^{i}_{CO}}+\frac{\epsilon^{i}_{CO}}{\epsilon^{i}_{CO}-\sigma^{i}_{CO}}}{2}\right)^{N_{b}(0,1)}\label{eq:cocoeff}	
\end{eqnarray}
where $\epsilon^{i}$ and $\sigma^{i}$ are the reported CO and CO$_2$ emission factors and their associated uncertainties for each sector $i$; $N_{a}(0,1)$ and $N_{b}(0,1)$ are $j^{\mathrm{th}}$ random numbers drawn from normal distributions with zero mean  and unity variance. Eqs \ref{eq:co2coeff} and \ref{eq:cocoeff} approximate log-normal uncertainties for emission factors $\epsilon$ is based on reported variability $\sigma$ (see Table~\ref{tab:emfac}) in order to avoid  unphysical (negative) values. 
 A thousand independent random  samples for $  r^{j}_{CO_2:CO}$ \add[KB]{are} taken from a normal distribution with standard deviation $\sigma^{i}_{CO_2}$, $\sigma^{i}_{CO}$ at each timestep and gridbox. The resulting second order moment of the distribution is approximated as 
 \begin{equation}
 E[(r_{CO_2:CO}-\overline{r_{CO_2:CO}})^2] \approx \mathnormal{Var}(r_{CO_2:CO})
 \label{eq:varco2co}
 \end{equation}
 where $\mathnormal{Var}(\cdot)$ is the empirical variance computed from Eqs.~\ref{eq:co2coeff} and~\ref{eq:cocoeff}.   
 
 The uncertainty in the biomass burning CO$_2$ is the product of the $r_{CO_{2}:CO}$ uncertainty and the uncertainty in the CO emissions.  The 2\textsuperscript{nd} order moments are \citep{Simon:2002aa}
 \begin{eqnarray}
 E[r_{CO_{2}:CO} \epsilon_{CO}] & = & \overline{r_{CO_{2}:CO}} \overline{\epsilon_{CO}} \\
 E[(r_{CO_{2}:CO} \epsilon_{CO}- \overline{r_{CO_{2}:CO}} \overline{\epsilon_{CO}})^2 ] &=& \overline{r_{CO_{2}:CO}}^2 \sigma^2_{CO}+ \overline{\epsilon_{CO}}^2 \sigma^2_{CO_{2}:CO} + \sigma^2_{CO_{2}:CO} \sigma^2_{CO} 
 \end{eqnarray}
 where $ \sigma^2_{CO_{2}:CO} $ is computed from Eq.~\ref{eq:varco2co} and $\sigma^2_{CO}$ is computed from a Monte Carlo simulation described in \cite{Liu:2014kx}.  
 
\subsection{GPP}
\label{sec:gpp}

%\begin{itemize}
%\item Description of Parazoo et al, results. 
%\item Include brief description of equations?  {\color{blue} Parazoo}
%\end{itemize}
 Gross Primary Productivity, which  is enabled at leaf level by photosynthesis, is the major driver of carbon accumulation in the global terrestrial ecosystem \citep{Beer:2010vn}. A number of approaches have been developed to estimate the spatial patterns and magnitude of Gross Primary Productivity (GPP) including statistical interpolation of eddy-covariance flux tower measurements, optical reflectance measurements, and terrestrical ecosystem models \citep{Anav:2015ly}.  GPP can be modeled as the product of incoming photosynthetically active radiation (PAR), the fraction absorbed by vegetation (fAPAR), and  Light Use Efficiency (LUE), which quantifies the fraction of absorbed radiation used by photosynthesis \citep{Monteith:1972aa}.   A relatively new remote-sensing approach to infer GPP is based upon measurements of solar-induced chlorophyll fluorescence (SIF) \citep{Frankenberg:2011zr,Frankenberg:2014aa,Guanter:2012aa,Joiner:2011gf,Joiner:2013aa}.  SIF is emitted at the leaf level in the \add[KB]{visible to near-infrared} (660--800 nm) in response to solar radiation and like GPP is proportional to PAR and fAPAR. Since photosynthesis through LUE and SIF both compete for the same energy absorbed through photosynthesis, SIF has the potential of being a more physically driven proxy of GPP and its response to climate variability \citep{Porcar-Castell:2014fk}. Satellite-based SIF observations have been used to examine the response of Amazonian productivity to stress \citep{Lee:2013aa} and the relationship to net ecosystem exchange \citep{Parazoo:2013fk}.  These applications have expanded to include crop productivity \citep{Guanter:2014uq} and ENSO response \citep{Parazoo:2015aa,Yoshida:2015ty}. While \add[KB]{knowledge of }the precise relationship between GPP and SIF is \add{continuing to evolve,} \remove[KB]{evolving} recent validation activities are promising \citep{Damm:2015yq,Yang:2015aa}.  The assimilation of SIF into terrestrial ecosystem models \change[KB]{are}{is} being explored \citep{Parazoo:2014fk,Zhang:2014aa,Koffi:2015qy}.

GPP estimates are integrated within CMS-Flux following \cite{Parazoo:2014fk}.   Based upon the same formalism in Eq.~\ref{eq:cf}, monthly GPP at each grid point is inferred from a precision-weighted minimization of GOSAT SIF, which is regressed against global GPP from upscaled flux tower data \citep[e.g.,][]{Frankenberg:2011mz,Jung:2011kx}  and is also subjected to a priori knowledge of GPP variability derived from an ensemble of terrestrial ecosystem models.  The approach here differs from \cite{Parazoo:2014fk} \change[KB]{where}{in that} the prior mean is taken from CASA-GFED (Sec. \ref{sec:te}).    

\section{Results}

\subsection{Global flux tendencies}
\label{sec:results}
\begin{figure}
		\centering
		\includegraphics[width=15cm]{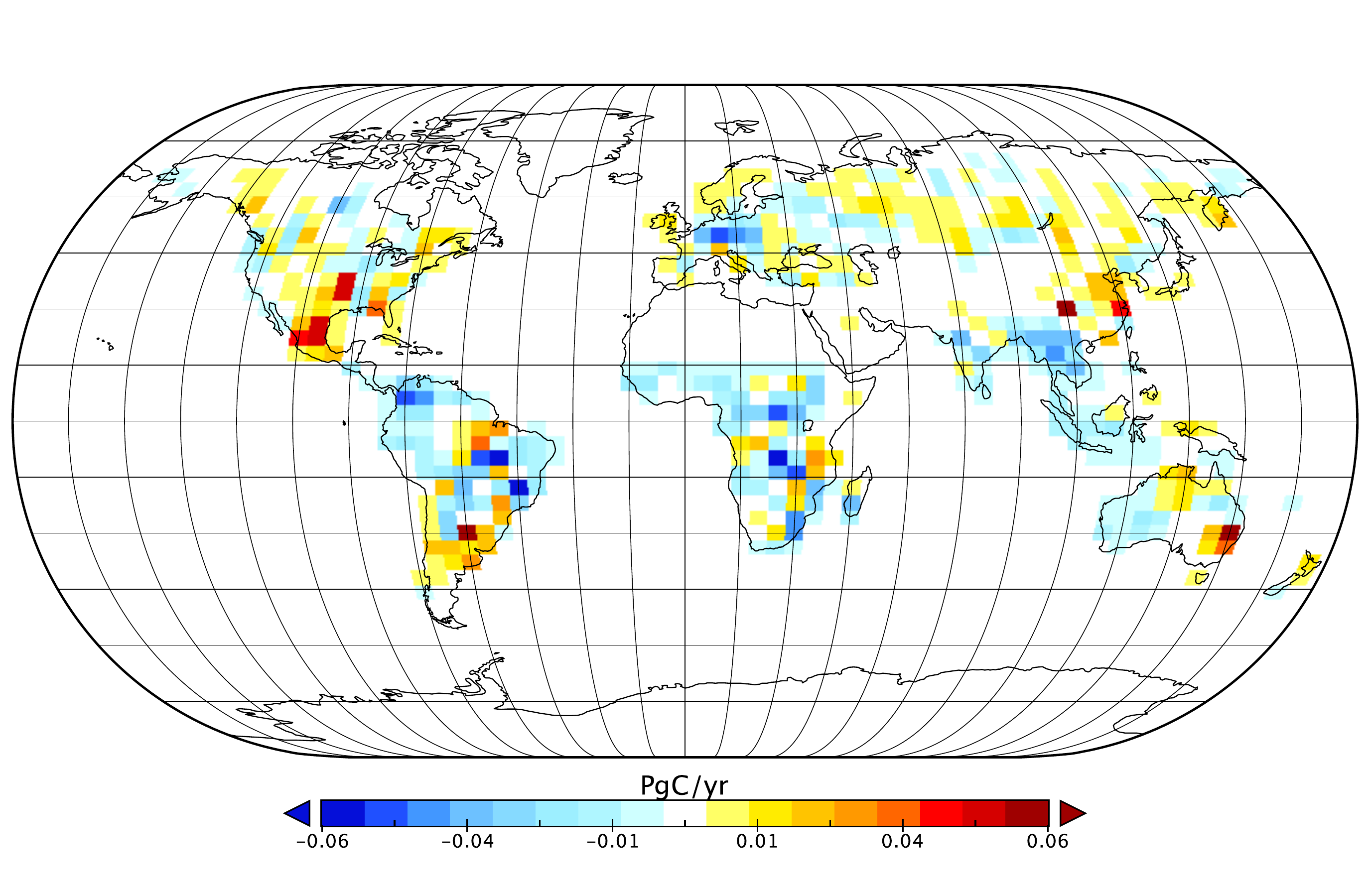}
		\caption{(a) Total CO$_2$ flux tendency from 2011-2010. }
		\label{fig:dfluxtotmapTotaleck}
	\end{figure}

%\begin{itemize}
%\item Show total map of change
%\item Discuss change in y-H(x) before and after flux estimate
%\item Note consistency with previous results: Poulter, Basu, Gatti, Parazoo 
%\item summary statistics of TCCON and surface differences?
%\end{itemize}
%TODO: Make table of summary statistics of TCCON and surface data
The net carbon flux tendency for 2011-2010 is shown at 4$^{\circ}\times5^{\circ}$ in Fig.~\ref{fig:dfluxtotmapTotaleck}. The residual mean error, $E[\mathbf{y}-{\cal H}(\mathbf{x})]$, before and after the assimilation is described in Table~\ref{tab:ymhx}.  The prior mean error is significantly less than 1 ppm because the prior is constructed to be consistent with the atmospheric growth rate. The posterior residual mean error is less than the prior for both the tropics and extratropics. In particular, the Northern Hemisphere (NH) posterior  residual mean error is  $\leq 0.1$ ppm for both 2010 and 2011.  For both years, the prior prediction overestimates  extra-tropical xCO$_2$  whereas the tropics is underestimated.   

The net carbon flux tendency is  -1.60 PgC, which tends to be positive in the Northern Hemisphere and negative in the tropics though with significant spatial variability.  The regional distributions are consistent with previously reported results; though these tend to focus on absolute flux. Southeast Asia fluxes show a strong negative tendency (-0.27 $\pm$ 0.05 PgC) consistent with the weak 2010 uptake reported in \cite{Basu:2013kx}.  Europe also shows a stronger uptake in 2011 (-0.16 $\pm$ 0.05 PgC) similar to previous reports \citep{Reuter:2014ab,Deng:2014fk}. The impact of La Ni\~{n}a results in higher flux in 2011 relative to 2010 in Mexico and Texas (0.25 $\pm$ 0.04 PgC) \citep{Parazoo:2015aa}.  The flux \change[KB]{tendencies}{tendency} in Brazil was -0.24 PgC and will be discussed in more detail in Sec.~\ref{sec:brazil}. 
%TODO Need to calculate those numbers. 
%TODO Need to find Feng results for Europe
%\begin{figure}
%\centering
%\includegraphics[width=0.7\linewidth]{Plots/dfluxtotmapTotaleck}
%\caption{Total CO$_2$ flux tendency from 2011-2010}
%\label{fig:dfluxtotmapTotaleck}
%\end{figure}

%\begin{figure}
%	\centering
%	\includegraphics[width=0.7\linewidth]{Plots/dfluxtotmapBB20112010}
%	\caption[BB]{Biomass burning flux tendency from 2011-2010 constrained by MOPITT CO columns}
%	\label{fig:dfluxtotmapBB20112010}
%\end{figure}

\remove[KB]{The El Ni\~{n}o/La Ni\~{n}a in 2010-2011 was the largest Central Pacific ENSO event in the satellite record. This led to a record-breaking tropical North Atlantic temperature anomaly that in turned caused the once-in-a-century Amazonian droughts \protect\citep{Lewis:2011uq}}. \add[KB]{As the El Ni\~{n}o Modoki is} a coupled atmosphere-ocean response, the ocean carbon response was also important.  The ocean carbon was estimated based upon the ECCO2-Darwin model (Sec.~\ref{sec:ecco2}) using the assimilation technique described in \cite{Brix:2015fk}.  The tendency for each ocean basin is listed in Table~\ref{tab:ocean}. The Atlantic ocean basin has the strongest tendency at -0.187 PgC.  The transition from El Ni\~{n}o to La Ni\~{n}a largely offset the Pacific ocean carbon response whereas the Atlantic ocean anomalous sea surface temperatures (SST) \change[KB]{where}{were} not. The Atlantic ocean tendency is approximately half of the total ocean tendency.

%\begin{itemize}
%	\item Show global ocean tendency map?
%	\item Or just show basin scale estimates
%	\item Calcluate what the TNA contribution is. 
%\end{itemize}

\begin{figure}
		\centering
		\includegraphics[width=15cm]{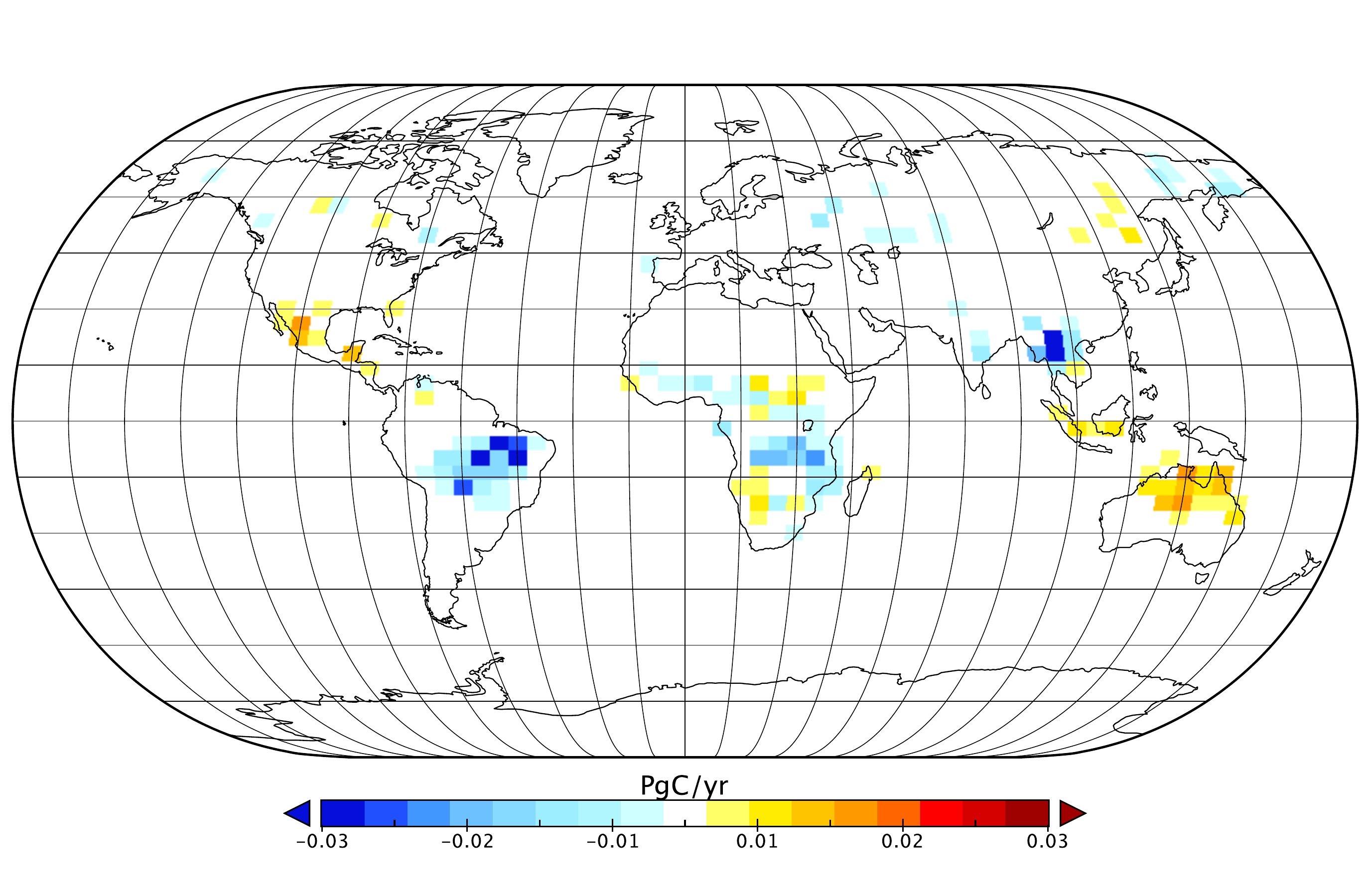}
		\caption{(b) Biomass burning flux tendency from 2011-2010 constrained by MOPITT CO columns.}
		\label{fig:dfluxtotmapBB20112010}
	\end{figure}

%\subsection{Biomass burning tendency}
%\label{sec:bbt}
%\begin{itemize}
%\item Show biomass burning map
%\item contrast brazilian flux against SE Asia--maybe reference Basu's results
%\item Discuss the biogenic component?  
%\end{itemize}
%TODO update biomass burning to Zhe's new results. 
%TODO compare results between column and surface
%TODO need to look at biogenic component in Brazil.
%TODO Calculate Southeast Asia and Africa; maybe add in table.  
The biomass burning tendency constrained by MOPITT using the \remove[KB]{the} approach discussed in Sec.~\ref{sec:bb} is shown in Fig.~\ref{fig:dfluxtotmapBB20112010}.  Overall, there was more carbon \remove[KB]{fluxes} emitted from biomass burning globally in 2010 than in 2011 with most of the changes occurring in the tropics, namely sub-equatorial Africa, South East Asia, and Brazil.  Of those, biomass burning in the ``arc of deforestation" in Brazil dominates the tendency accounting for approximately 50\% of the global \add[KB]{biomass burning tendency}.  While African fires contribute significantly to  annual biomass burning, the 2011 burning was only slightly lower than 2010.   The biomass burning tendency was important in Southeast Asia in 2010 (0.08 $\pm$ 0.001 PgC), which was also observed with the tropospheric CO from the Infrared Atmospheric Sounding Interferometer (IASI)~\citep{Basu:2014fk}, and represented about 30\% of the net carbon flux tendency.   On the other hand, subtropical Australia had larger biomass burning in 2011 (0.13 $\pm$ 0.01 PgC), likely a consequence of the La Ni\~{n}a.   The magnitude of this burning is determined in part by the available fuel load, which will increase during more productive and wet years \citep{Randerson:2005dq}. 
%\begin{figure}
%\centering
%\includegraphics[width=0.7\linewidth]{Plots/dfluxtotmapBB20112010}
%\caption[BB]{Biomass burning flux tendency from 2011-2010 constrained by MOPITT CO columns}
%\label{fig:dfluxtotmapBB20112010}
%\end{figure}

\begin{figure}
		\centering
		\includegraphics[width=15cm]{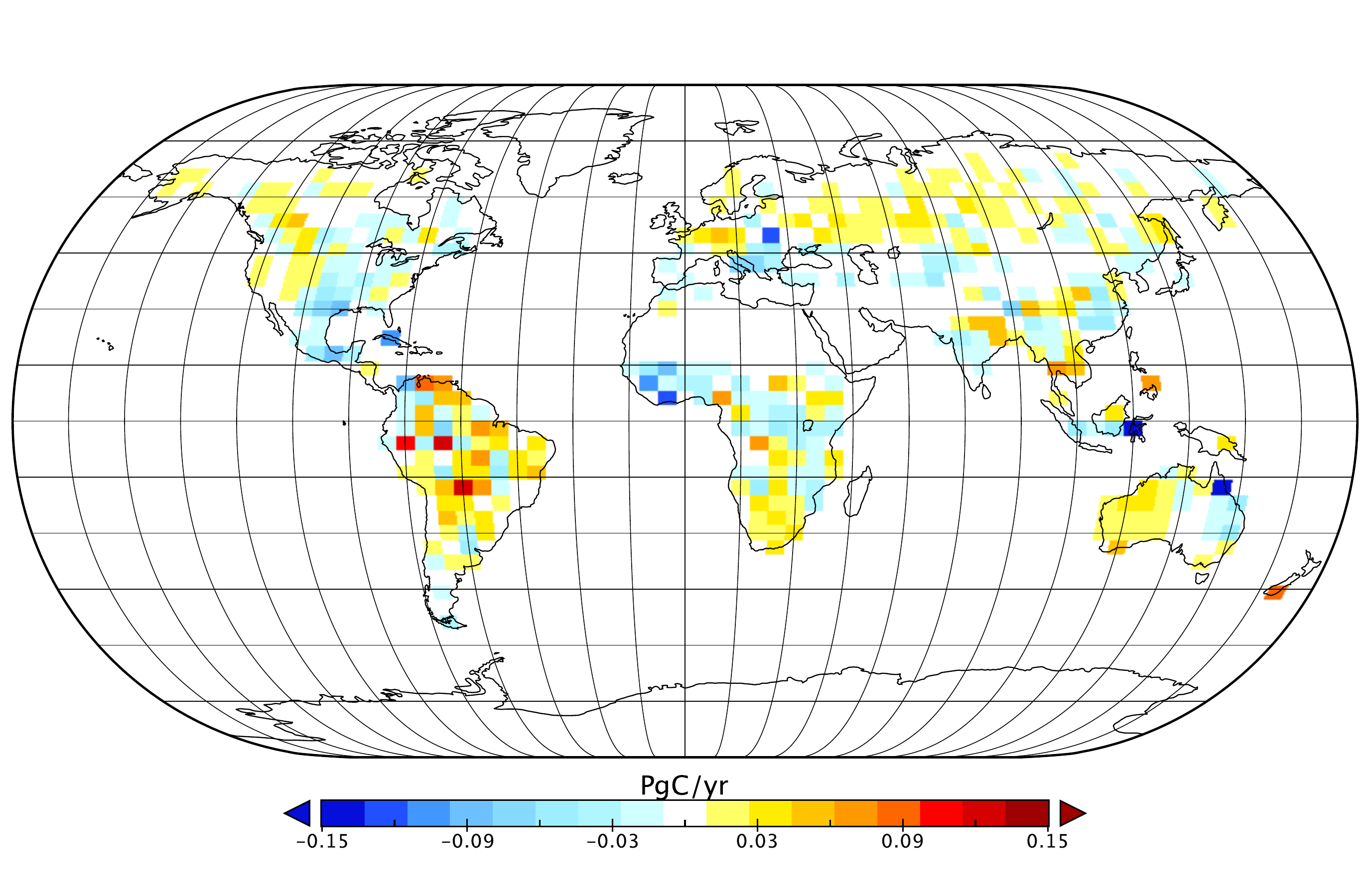}
		\caption{(c) The GPP tendency 2011-2010 based upon a combination of SIF from GOSAT  and terrestrial ecosystem models. }
		\label{fig:dcmsflux2011-2010GPP}
	\end{figure}

%\subsection{GPP tendency}
%\begin{itemize}
%\item Point at that Brazil is noisy because of cloud cover
%\item Point to gradient across Australia
%\item Note opposite response in Texas and Mexico
%\end{itemize}
The GPP tendency for 2011-2010 following the approach outlined in Sec.~\ref{sec:gpp} is shown in Fig.~\ref{fig:dcmsflux2011-2010GPP} where the color scale is reversed for comparison with the net carbon flux in Fig.~\ref{fig:dfluxtotmapTotaleck}.  In the southern midlatitudes, the GPP tendency is higher in western Australia (0.13 $\pm$ 0.06 PgC) consistent with a greater uptake in the net carbon flux in Fig.~\ref{fig:dfluxtotmapTotaleck}.  Eastern Australia had the opposite tendency. The total regional flux  has a more complex structure, which may reflect the combination of natural and anthropogenic sources, including the significant increases in biomass burning in Northern Australia (Fig.~\ref{fig:dfluxtotmapBB20112010}).  The GPP tendency is not uniformly positive across Southeast Asia, which is approximately neutral at 0.08 $\pm$ 0.30 PgC,  whereas the biomass burning was sharply reduced over specific regions suggesting that the driver of the Southeast Asia net carbon flux tendency is primarily through a combination of burning and respiration processes.  

Mexico and Texas have a significantly reduced GPP tendency, which is consistent with a reduced uptake in net carbon flux.  This change reflects the impact of the strong La Ni\~{n}a in 2011 \citep{Parazoo:2015aa}.  

%\begin{figure}
%\centering
%\includegraphics[width=0.7\linewidth]{Plots/dcmsflux2011-2010GPP}
%\caption[GPP]{The GPP tendency 2011-2010 based upon a combination of GOSAT and terrestrial ecosystem models.}
%\label{fig:dcmsflux2011-2010GPP}
%\end{figure}

\subsection{Brazilian carbon tendency in global context}
\label{sec:brazil}
%\begin{itemize}
%\item Evaluate net carbon flux against Gatti aircraft measurements 
%\item Describe differences between the different regions in terms of the Liu and Bowman paper. 
%\item discussion of source/receptor relationship between Amazonian fluxes and SH observations, citing Liu, Bowman, and Henze et al. Include discussion of cloud cover. 
%\item Show flux decomposition bar chart.  
%\item according to flux estimate, Brazilian carbon tendency is explained entirely by biomass burning
%\item By residual analysis, NEP had to be nearly zero. 
%\item Given that GPP was positive, an NEP of zero implies that GPP was balance by increased respiration.  
%\item Note that 2011 had very high precipitation, which could stimulate respiration processes. 
%\item implied respiration can not attribute the respiration to a specific carbon pool.
%\item could have a potential impact on mortality, but additional years are necessary.
%\item Brazilian carbon tendency was about 10\% of total tendency, which was about 1.5PgC as seen by GOSAT.  
%\begin{itemize}
%\item Should note from Liu et al, 2014, that there is a sampling bias with NIR satellites that do not capture the full seasonal cycle. 
%\end{itemize}
%\item NEP was nearly zero for Brazil but was the dominant factor in the net carbon flux tendency.
%\item However, biomass burning was about 25\% of the land.  Half of that BB was attributable to Brazil.   
%\end{itemize}
\add[KB]{Applying Eq.~\protect{\ref{eq:dflux}}, the  flux tendencies for Brazil are shown in Fig.~\ref{fig:gdbrazil20112010uncbar} incorporating the independent estimates of net carbon flux, GPP, and biomass burning inferred from GOSAT and MOPITT observations.    The fossil fuel, ocean, and chemical source tendency are  negligible based upon the bottom up estimates.  The contribution of riverine carbon flux is implicitly folded into the terrestrial and ocean carbon budgets because atmospheric data cannot readily partition and attribute lateral fluxes. From these considerations,  the NEP and its uncertainty is calculated as a residual of the net carbon flux tendency and biomass burning.  

The net carbon flux tendency of $-0.24\pm 0.11$ PgC  can be attributed almost entirely to the biomass burning tendency, which are in the range  of -0.32 PgC from land data and -0.15 PgC from land+ocean data in \cite{Deng:2016aa}. Consequently, the NEP tendency must be neutral, which is surprising given the strong drought in 2010.  The GPP tendency for Brazil is 0.5 PgC confirming that 2011 had higher productivity as expected.  The respiration tendency is computed as the final residual from substituting Eq.~\ref{eq:NEP} into Eq.~\ref{eq:dflux}.   The neutral NEP implies that the positive GPP tendency is balanced by the respiration tendency.  

%I'm calculating the tendency uncertainties assuming independent uncertainties for each year 
The net carbon flux tendency is broadly consistent with previous estimates that primarily use aircraft observations.  For the Amazonian carbon tendency,  \cite{Laan-Luijkx:2015aa} estimated an annual change in net carbon flux between  -0.24 and -0.50 PgC/yr  for  2011 relative to 2010 using a range of top-down and bottom-up estimates.  Based upon the assimilation of aircraft and thermal infrared satellite measurements, the net carbon flux tendency was -0.34$\pm$0.60 PgC yr$^{-1}$, which is slightly higher than our results, but   less than the -0.42$\pm$0.2 PgC yr$^{-1}$ from \cite{Gatti:2014uq}.   The fire emission tendency for \cite{Laan-Luijkx:2015aa}, \cite{Gatti:2014uq} and this study are between -0.21 and -0.24 PgC yr$^{-1}$.  \cite{Alden:2016aa}  used the aircraft in \cite{Gatti:2014uq} to estimate a basin-wide net carbon flux of 0.28$\pm$0.45 PgC.  Based upon the uncertainties, however, that study can not reject a neutral biosphere tendency. The increase in 2011 relative to 2010 of both respiration and  GPP could be explained by a number of processes, such as a link between heterotrophic respiration and soil moisture \citep[e.g.,][]{Exbrayat:2013kq}, or the lagged effects such as tree mortality, \citep[e.g.,][]{Saatchi:2013uq} that would offset the increased GPP from faster carbon pools.  On the other hand, \cite{Doughty:2015aa} estimated a relatively constant NPP of 0.38 PgC (0.22-0.55 PgC) based upon upscaled forest plot data for both years. This difference would imply a more spatially heterogeneous response than what is captured by site data.

An important consideration in comparing these studies is the relationship between the observational coverage and the spatial domain of fluxes that influences those observations.  In the case of flux inversions using aircraft data, the zonal CO$_2$ gradient across the basin was exploited. }
%It is therefore possible that the drought increased available carbon through litter fall or possibly tree mortality, which was associated with a previous drought in 2005 \citep{Saatchi:2013uq}. The precipitation tendency from the Global Precipitation Climatology Project (GPCP) over Brazil was on average 26 mm month$^{-1}$, which could have stimulated respiration processes.  On the other hand, 2011 was also cooler with an average MERRA temperature anomaly of approximately -0.5K. However,  a two year analysis is insufficient to attribute the change in fluxes to changes in carbon pools.
\remove[KB]{Estimating changes in carbon flux in Brazil is especially challenging because of pervasive cloud over the Amazon, especially in the west.} \add[KB]{Flux inversions with satellite column CO2 observations, on the other hand, use observations over a much broader domain based on the sensitivity of the observations to surface fluxes. }{\cite{Liu:2015ys}} quantified the source-receptor relationships between concentrations and fluxes for January and July.  This study showed a strong impact of tropical S. American fluxes on midlatitude observations with sensitivities exceeding 0.2 ppm/KgC/m$^2$/s (see Fig. 6 in \cite{Liu:2015ys}). The transit time from  concentrations emitted  in tropical S. America to midlatitude S. America is within a couple of days and the dwell time lasts over a month indicating continual influence of tropical fluxes.  Based upon an Observing System Simulation Experiment,  removal of midlatitude S. American observations led to over a 50\% impact on western Amazonian fluxes where cloud cover is most persistent (See Fig. 12 in \cite{Liu:2015ys}).  On the other hand, tropical S. American concentrations, especially eastern Amazon, are influenced by tropical African fluxes in both January and July.  The cumulative sensitivity over a 1-month time frame of tropical African fluxes to tropical S. American concentrations is roughly 25\% (see Fig. 12 in \cite{Liu:2015ys}). \add[KB]{Consequently, the inversion system exploits the meridional source-receptor relationships between Amazonian fluxes and midlatitude GOSAT observations more so than basin-wide zonal gradients observed  by aircraft. }

These source-receptor relationships facilitate the interpretation of posterior CO$_2$ concentration comparisons to independent aircraft observations used in \cite{Gatti:2014uq}. Aircraft spirals were taken twice a week from 2010-2011 over   Rio Branco (RBA), Tabatinga (TAB),  Alta Floresta (ALF) and Santar\'{e}m (SAN) as shown in the top panel in Fig.~\ref{fig:aircraft}. The root-mean-square (RMS) difference between \add[KB]{aircraft and} CMS-Flux prior (blue) and posterior (red) \add[KB]{model concentrations for both years} are shown in the bottom panel.  The best agreement is at TAB in Western Amazon with about a 1 ppm RMS error above 2km. This \change[KB]{approved}{improved} agreement is likely a consequence of the strong sensitivity of this region to midlatitude GOSAT observations.   \change[KB]{However, there is a persistent positive bias below 1km up to 5 ppm, which suggest that the absolute fluxes are high biased. Both RBA and ALF show improved agreement but still positively biased. On the other hand, SAN posterior RMS error changes very little relative to  the prior.}{However, the RMS is persistently high below 1km ($\approx$ 5 ppm).  Near surface CO2 concentrations are influenced by boundary level dynamics and local flux forcing that would be difficult to simulate with a global scale analysis. Both RBA and ALF show improved agreement in RMS by up to 1 ppm relative to the prior.  These improvements suggest that the transport model forced by posterior fluxes are better able to simulate the aircraft CO$_2$ variability. On the other hand, SAN posterior RMS error changes very little relative to  the prior. }  The eastern Amazon is more strongly affected by tropical African fluxes through Atlantic cyclones than the other locations.  The assimilation system must attribute xCO$_2$ variability over tropical S. America and Africa to both local and non-local fluxes. The transport patterns over Africa are complex based upon the seasonal shift of the ITCZ and the African monsoon that may lead to model error, which partly explains the reduced agreement.  Nevertheless, the free tropospheric \add[KB]{RMS}  agreement is approximately 1 ppm.  

\remove[KB]{Improvement of posterior concentrations to independent observations indicates that some distribution of posterior fluxes should be more accurate than prior fluxes.  \protect\cite{Liu:2016aa} introduced a new methodology to attribute improved agreement of independent concentration data to the accuracy of inferred fluxes. A cost function is constructed from the squared difference between the aircraft and predicted CO$_2$ for prior and posterior concentrations separately leading to $J_{prior}$ and $J_{post}$ (see Eq. 1 in \cite{Liu:2016aa}). The difference $J_{post}-J_{prior}$ is 1018, -785, -2037, -4035 ppm$^2$ for SAN, TAB, ALF, and RBA respectively calculated over a two year period. The negative values for TAB, ALF, and RBA indicate improved agreement between posterior CO$_2$ and aircraft observations, i.e., $J_{post} < J_{prior}$.  The contribution of fluxes to this difference for each aircraft site is shown in Fig.~\ref{fig:Reduction_Amazonia_random_rms-clipv2}. The 3 sites that have improved agreement are driven by fluxes primarily to the west of the aircraft sites whereas at SAN, which has reduced agreement, is driven by fluxes northeastern Brazil and Africa.  Based upon this approach, posterior fluxes in western Amazon are more accurate than fluxes in the northeast of Brazil. }
\add[KB]{Improvement of posterior concentrations to independent observations indicates that some distribution of posterior fluxes should be more accurate than prior fluxes.  \protect\cite{Liu:2016aa} introduced a new methodology to attribute improved agreement of independent concentration data to the accuracy of inferred fluxes. Two cost functions are constructed from the vertically summed squared difference between the aircraft and predicted CO$_2$ for prior and posterior concentrations:
\begin{eqnarray}
J_{post}(\mathbf{x}) &=& (\mathbf{c}_{\mathrm{aircraft}}-{\cal H}({\bf{x}}_{\mathrm{post}}))^{\top}(\mathbf{c}_{\mathrm{aircraft}}-{\cal H}({\bf{x}}_{\mathrm{post}})) \\
J_{prior}(\mathbf{x}) &=& (\mathbf{c}_{\mathrm{aircraft}}-{\cal H}({\bf{x}}_{\mathrm{prior}}))^{\top}(\mathbf{c}_{\mathrm{aircraft}}-{\cal H}({\bf{x}}_{\mathrm{prior}}))	
\end{eqnarray}
The difference $J_{post}-J_{prior}$ is 931, -985, -669, -3366 ppm$^2$ for SAN, TAB, ALF, and RBA respectively calculated over a two year period. Consistent with Fig.~\ref{fig:aircraft}, the negative values for TAB, ALF, and RBA indicate improved agreement between posterior CO$_2$ and aircraft observations, i.e., $J_{post} < J_{prior}$. The sensitivity of the cost function difference to fluxes can be calculated with an adjoint \citep{Liu:2016aa}.  Based upon this approach, the contribution of fluxes to this difference for each aircraft site is shown in Fig.~\ref{fig:Reduction_Amazonia_random_rms-clipv2}. Negative values indicate the adjustment to the prior fluxes during flux inversion improves agreement with aircraft whereas positive values indicate the adjustment reduces agreement.  The CO$_2$ at the 3 sites that have improved agreement are sensitive to fluxes primarily to the west of the aircraft sites whereas CO$_2$ at SAN, which has reduced agreement, is sensitive to fluxes in northeastern Brazil and Africa.  Based upon this approach, posterior fluxes in western Amazon are more accurate than the prior fluxes, while the posterior fluxes in the northeast of Brazil are less accurate. None of the sites, however, are sensitive to Brazilian fluxes south of the Amazonian basin.}

\remove[KB]{This analysis illustrates the complexities in comparing fluxes derived from aircraft and satellite observations as well as combining multiple measurements to attribute net fluxes to gross fluxes. Our study is consistent with  previously discussed studies in that total and fire fluxes where higher in 2010 relative to 2011.  As discussed in \protect\cite{Laan-Luijkx:2015aa},  regionally integrated NEP fluxes are a sensitive balance beween total and fire fluxes, which are spatially heterogeneous. It is therefore possible that the drought increased available carbon through litter fall or possibly tree mortality, which was associated with a previous drought in 2005 \citep{Saatchi:2013uq}. The precipitation tendency from the Global Precipitation Climatology Project (GPCP) over Brazil was on average 26 mm month$^{-1}$, which could have stimulated respiration processes.  On the other hand, 2011 was also cooler with an average MERRA temperature anomaly of approximately -0.5K. However,  a two year analysis is insufficient to attribute the change in fluxes to changes in carbon pools.   Consequently, the  interpretation of component flux tendencies in Fig.~\ref{fig:gdbrazil20112010uncbar}  in terms of specific ecosystem processes will be require a longer time period and additional analysis. }
  
\remove[KB]{Applying Eq.~\protect{\ref{eq:dflux}}, the  flux tendencies for Brazil are shown in Fig.~\ref{fig:gdbrazil20112010uncbar} incorporating the independent estimates of net carbon flux, GPP, and biomass burning inferred from GOSAT and MOPITT observations.    The fossil fuel, ocean, and chemical source tendency are  negligible based upon the bottom up estimates.  The contribution of riverine carbon flux is implicitly folded into the terrestrial and ocean carbon budgets because atmospheric data can not readily partition and attribute lateral fluxes. From these considerations,  the NEP and its uncertainty is calculated as a residual of the net carbon flux tendency and biomass burning.  

The net carbon flux tendency of $-0.24\pm 0.11$ PgC  can be attributed almost entirely to the biomass burning tendency, which is in the range  of -0.32 PgC from land data and -0.15 PgC from land+ocean data in \cite{Deng:2016aa}. Consequently, the NEP tendency must be neutral, which is surprising given the strong drought in 2010.  The GPP tendency for Brazil is 0.5 PgC confirming that 2011 had higher productivity as expected.  The respiration tendency is computed as the final residual from substituting Eq.~\ref{eq:NEP} into Eq.~\ref{eq:dflux}.   The neutral NEP implies that the positive GPP tendency is balanced by the respiration tendency, which is the sum of both autotrophic and heterotrophic respiration.  It is therefore possible that the drought increased available carbon through litter fall or possibly tree mortality, which was associated with a previous drought in 2005 \citep{Saatchi:2013uq}. The precipitation tendency from the Global Precipitation Climatology Project (GPCP) over Brazil was on average 26 mm month$^{-1}$, which could have stimulated respiration processes.  On the other hand, 2011 was also cooler with an average MERRA temperature anomaly of approximately -0.5K. However,  a two year analysis is insufficient to attribute the change in fluxes to changes in carbon pools.  }
\section{Conclusions}
  %% \conclusions[modified heading if necessary]
%\begin{itemize}
%\item Historic El Modoki ENSO event had a complex ocean-land impact on the carbon tendency (balance) in 2010-2011.  
%\item  Led to devastating droughts in Brazil
%\item ENSO is a Pacific ocean driven event but net impact was in the Atlantic ocean
%\item  complex carbon cycle response to ENSO indicates opposing responses, e.g., productivity and respiration, can mute regional responses.
%\item  high productivity in some regions can balance high respiration in other regions when looking at interannual variability. 
%\item  critical to disentangle BB, ocean, fossil fuel, productivity, and respiration processes. 
%\item  discuss implications to emergent constraints?  
%\end{itemize}
\label{sec:conc}
The El Ni\~{n}o Modoki  event has a complex land-ocean impact on the carbon tendency in 2010-2011.  While ENSO is primarily a Pacific process, the net carbon impact was centered primarily in the Atlantic with a net carbon tendency of -0.33 PgC.   The impact of Brazil of  -0.24 PgC  is only about 15\% of the net carbon flux tendency of  -1.6 PgC observed by GOSAT.  On the other hand, the global biomass burning tendency was -0.27 PgC, which is on par with the Brazilian tendency. From that standpoint, combustion processes in Brazil played a dominant role in the biomass burning component of the global carbon cycle.   However, the dominant driver of the global carbon tendency was NEP at 1.26 PgC.  

The interpretation of the Brazilian drought in the global carbon tendency from 2011-2010 is further complicated by the La Ni\~{n}a in 2011, which led to strong precipitation in Australia~\citep{Fasullo:2013aa} but strong droughts in Texas and Mexico~\citep{Parazoo:2015aa}. The response of GPP and respiration to those regional drivers will vary depending on the biome such as semi-arid regions where the interannual variability, for example,  has been attributed to GPP~\citep{Poulter:2014fk}.  For regions such as the Amazon, the distribution of carbon pools makes attribution of interannual variability  more difficult as these pools have different time scales and responses to climate drivers \citep{Carvalhais:2014aa,Bloom:2016aa}.  The coarse spatial resolution of the flux estimate can not resolve biome distributions. Consequently, the increase in GPP in 2011 could be attributable to a combination of the older forests and regrowth. Tree mortality \add[KB]{or interactions between respiration and soil moisture} as a consequence of the drought can not be ruled out based upon the inferred respiration response \remove[KB]{and is consistent with previous studies} \citep{Gatti:2014uq,Brando:2014ys,Doughty:2015aa}.   

To fully quantify the response of the carbon cycle to climate variability and forcing, it is critical to disentangle the constituent processes and how they constructively or destructively interfere to drive the net atmospheric growth rate. The launch of the Orbital Carbon Observatories (OCO-2 and OCO-3) and Sentinel 5p (TROPOMI) will continue to provide CO$_2$, CO, and SIF observations needed to assess the longer term impacts of climate forcing \citep{Crisp:2004vn,Crisp:2012uq,Veefkind:2012aa}.  These observations along with observations needed to drive anthropogenic, oceanic, and terrestrial carbon models should help quantify the spatial drivers of interannual CO$_2$ variability and its dependence on climate variability including tropical land temperatures \citep{Wang:2013kx,Wang:2014aa,Cox:2013fk,Wenzel:2014ly}. 

%
%
%%\appendix
%%\section{\\ \\ \hspace*{-7mm} HEADING}    %% Appendix A
%%
%%\subsection                               %% Appendix A1, A2, etc.
%
%
%
%
\begin{acknowledgements}
 This research was carried out at the Jet Propulsion Laboratory, California Institute of Technology, under a contract with the National Aeronautics and Space Administration. We acknowledge support from the NASA Carbon Monitoring System (NNH14ZDA001N-CMS).  All computations were performed in NASA AMES supercomputers. GOSAT data is available at https://co2.jpl.nasa.gov/ and MOPITT data can be accessed from https://www2.acom.ucar.edu/mopitt. CMS-Flux results can be acquired at http://cmsflux.jpl.nasa.gov/. TCCON data were obtained from the TCCON Data Archive, hosted by the Carbon Dioxide Information Analysis Center (CDIAC) - tccon.onrl.gov. We acknowledge helpful discussions with Dr. John Miller and support in using the aircraft data. Funding for the aircraft data taken in Brazil were provided by UK NERC, and were downloaded from NCAS British Atmospheric Data Centre at \protect{http://catalogue.ceda.ac.uk/uuid/7201536a8b7a1a96de584e9b746acee3}.
\end{acknowledgements}

\bibliographystyle{agufull08} 
%\bibliography{Prelim_CMS_Flux_Rev}
%
\end{article}

%\begin{thebibliography}{}
%
%\bibitem[AUTHOR(YEAR)]{LABEL}
%REFERENCE 1
%
%\bibitem[AUTHOR(YEAR)]{LABEL}
%REFERENCE 2
%
%...
%
%\end{thebibliography}

%% Literature citations
%% command                        & example result
%% \citet{jones90}|               & Jones et al.\ (1990)
%% \citep{jones90}|               & (Jones et al., 1990)
%% \citep{jones90,jones93}|       & (Jones et al., 1990, 1993)
%% \citep[p.~32]{jones90}|        & (Jones et al., 1990, p.~32)
%% \citep[e.g.,][]{jones90}|      & (e.g., Jones et al., 1990)
%% \citep[e.g.,][p.~32]{jones90}| & (e.g., Jones et al., 1990, p.~32)
%% \citeauthor{jones90}|          & Jones et al.
%% \citeyear{jones90}|            & 1990

%% FIGURES %%%%%%%%%%%%%%%%%%%%%%%%%%%%%%%%%%%%%%%%%%%%%%%%%%%%%%%%%%%%%%%%%%%%

%% ONE-COLUMN FIGURES

	\begin{figure}
		\centering
		\includegraphics[width=15cm]{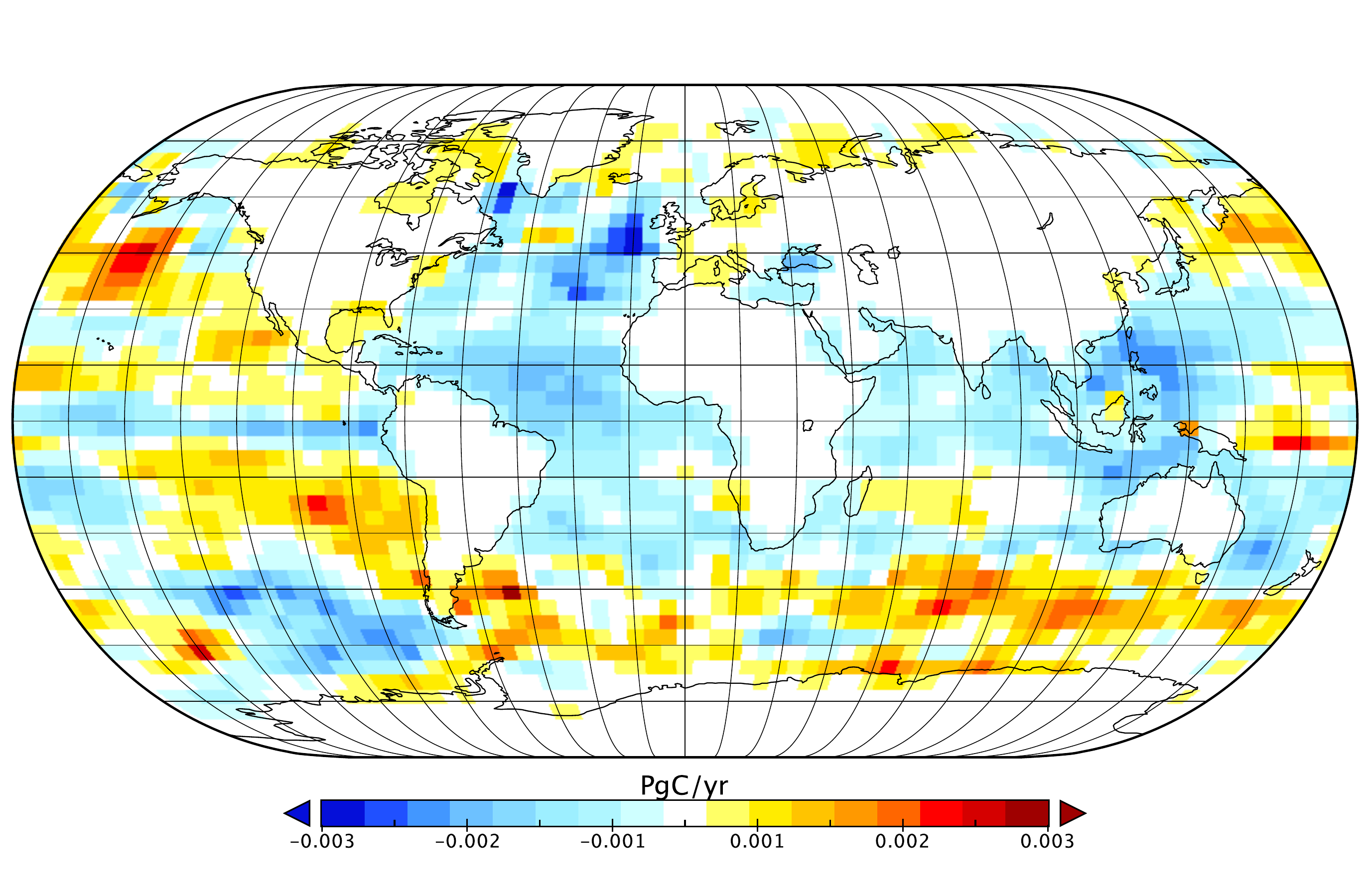}
		\caption{(d) The  ocean tendency from 2011-2010 based upon ECCO2-Darwin.}
		\label{fig:dcmsflux2011-2010OC}
	\end{figure}

\begin{figure}
	\centering
	\includegraphics[width=11cm]{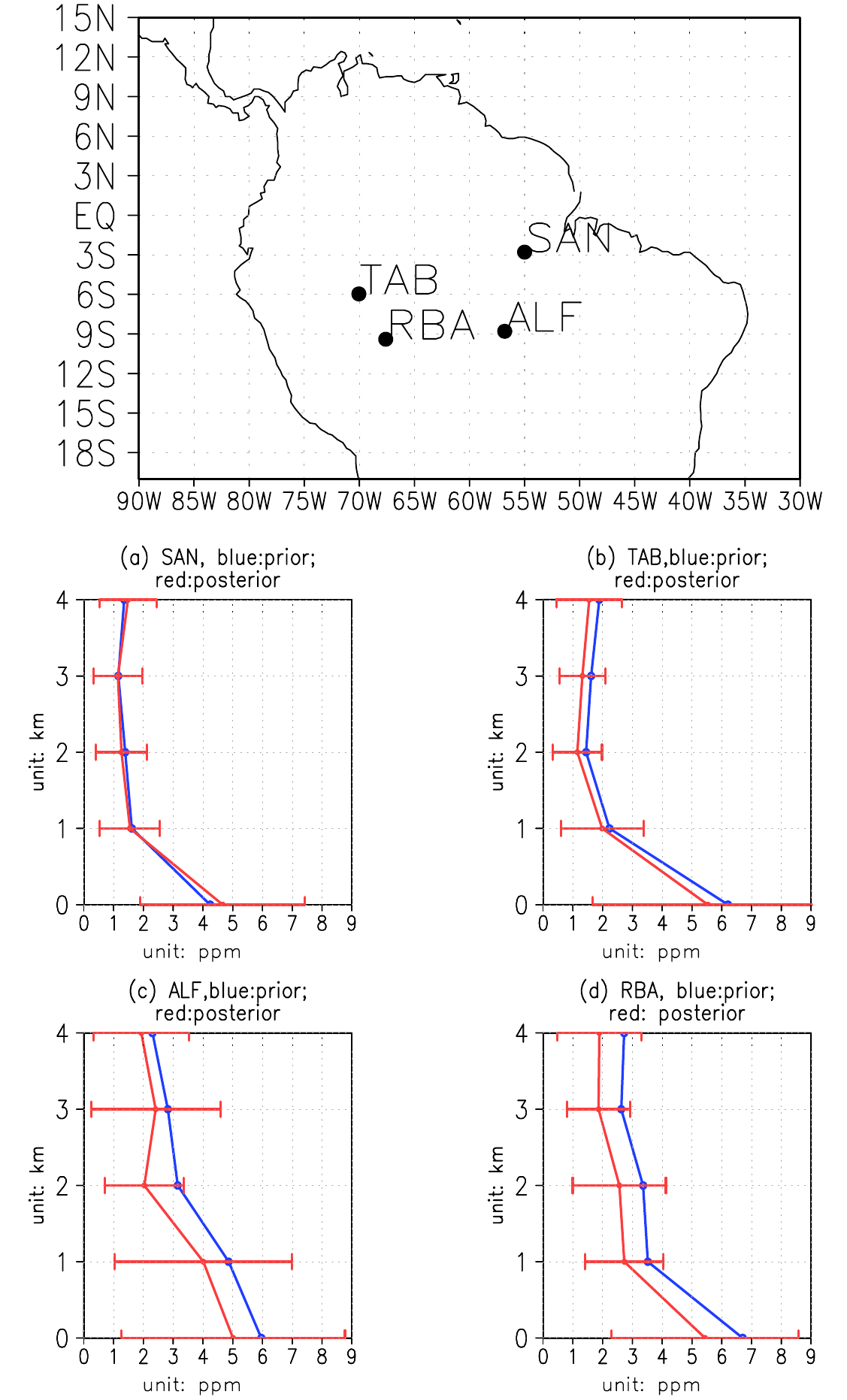}
	\caption{Location of aircraft observations used in \cite{Gatti:2014uq}. Aircraft spirals were taken twice per month 2010-2011. Bottom panels show the two-year mean RMS error between prior (blue) and posterior (red) CMS-Flux concentrations with respect to  aircraft observations. Site acronyms are Rio Branco (RBA), Tabatinga (TAB),  Alta Floresta (ALF) and Santar\'{e}m (SAN).  The standard deviation of the RMS error bars are shown  for the posterior concentrations.    }
	\label{fig:aircraft}
\end{figure}

\begin{figure}
	\centering
	\includegraphics[width=15cm]{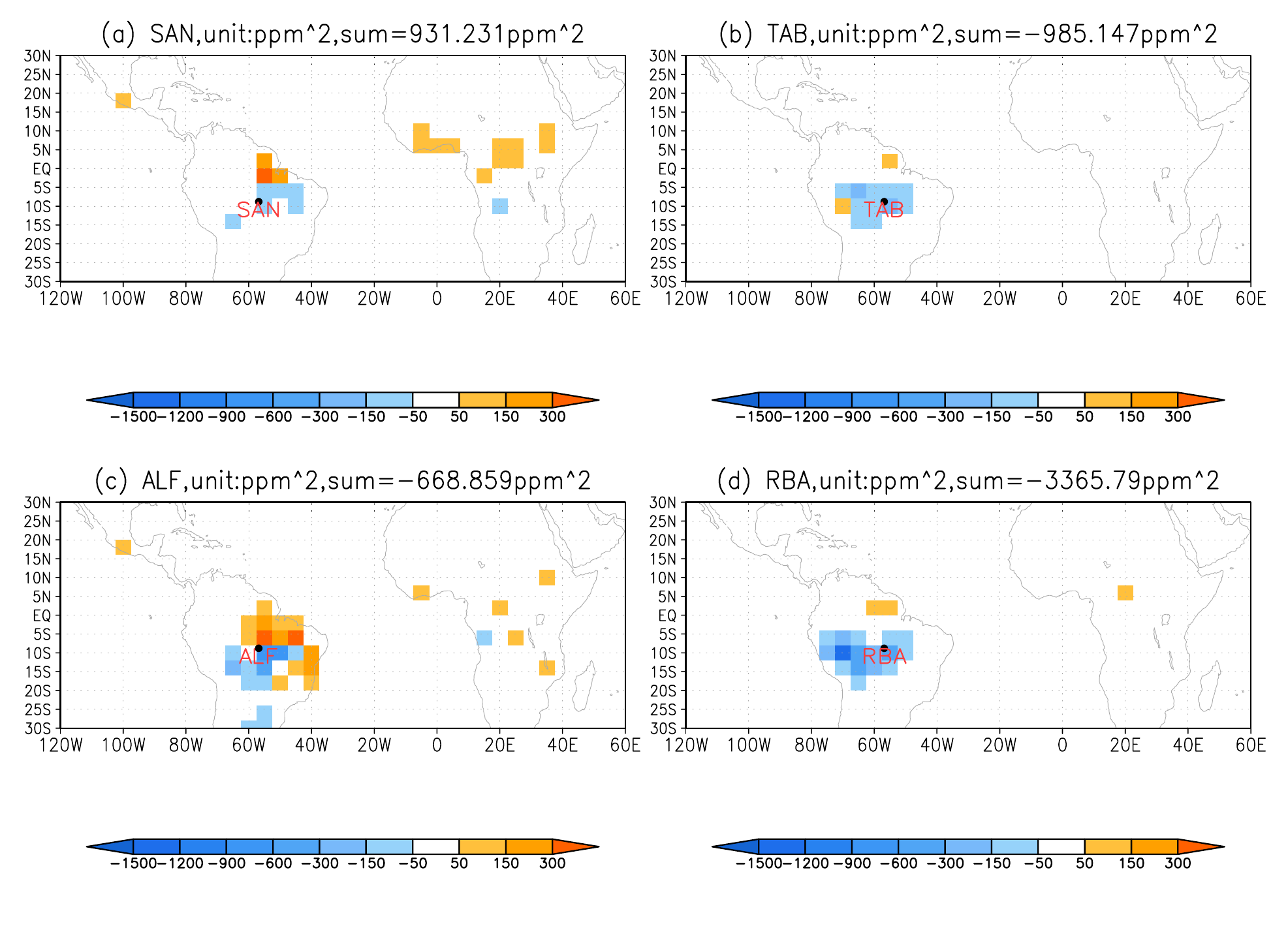}
	\caption{Distribution of fluxes impacting the validation of CMS-Flux CO$_2$ concentrations with respect to independent aircraft observations in Brazil. Negative values indicate fluxes that improve agreement with independent observations. The total change in agreement is denoted by the sum in the title for (a) Santar\'{e}m (SAN) (b) Tabatinga (TAB) (c) Alta Floresta (ALF) (d) Rio Branco (RBA).  }
	\label{fig:Reduction_Amazonia_random_rms-clipv2}
\end{figure}

\begin{figure}
	\centering
	\includegraphics[width=15cm]{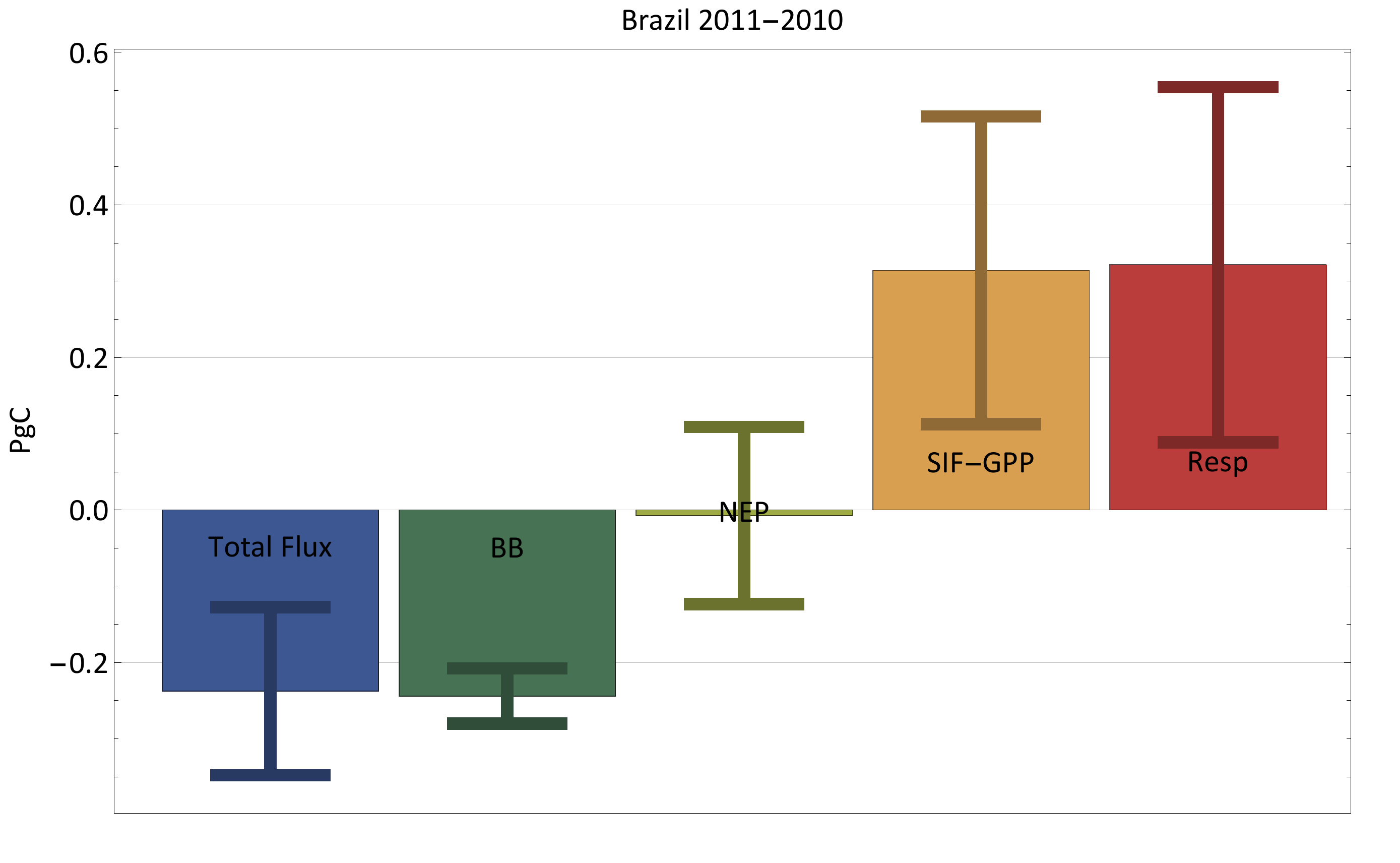}
	\caption[Brazil]{Flux tendency decomposed into total, biomass burning, NEP, GPP, and Respiration from 2010-2010. The net carbon flux and GPP are inferred independently from GOSAT v3.5 xCO$_2$ and SIF, biomass burning is inferred from MOPITT CO. NEP is computed as a residual of net carbon flux and biomass burning. Respiration is computed as a residual of NEP and GPP. }
	\label{fig:gdbrazil20112010uncbar}
\end{figure}

%% TABLES %%%%%%%%%%%%%%%%%%%%%%%%%%%%%%%%%%%%%%%%%%%%%%%%%%%%%%%%%%%%%%%%%%%%

%% ONE-COLUMN TABLE

%t
%\begin{table}[t]
%\caption{TEXT}
%\vskip4mm
%\centering
%\begin{tabular}{column = lcr}
%\tophline
%
%\middlehline
%
%\bottomhline
%\end{tabular}
%\end{table}

\begin{table}
	
	\begin{center}
		\caption{\label{tab:ymhx} Annual residual mean error between observations and model prediction ($\mathbf{y}-H(\mathbf{x})$) in ppm.}		
		\begin{tabular}{|ccccc|}\hline
			& 2010 &  & 2011  &  \\ \hline 
			Region	& Prior & Post & Prior  & Post \\ 
			15N--90N	& -0.24 & 0.03  & -0.11 & 0.08 \\ 
			15S--15N	& 0.34 & 0.31 & 0.42 & 0.26 \\ 
			15S--90S  	& -0.59  & -0.16 & -0.69 & -0.10 \\ \hline
		\end{tabular} 			
	\end{center}
	
\end{table}

\begin{table}[t]
\begin{center}
\caption{CMS-Flux global carbon budget for 2010 and 2011. }
\vskip4mm
\begin{tabular}{|c|c|c|c|} 
\hline
Year   & 2010 & 2011 & uncertainty \\ \hline
Growth rate (ppm) & 2.34 & 1.84 & 0.11 \\ \hline
Fossil Fuel (PgC) & 9.25 & 9.50 & 0.5  \\ \hline
Ocean (PgC)  & 2.38 & 2.71 & 0.5 \\ \hline
Fire (PgC) & 2.92 & 2.24 & ? \\ \hline
Terrestrial (PgC) & 4.85 & 5.79 & 1 \\ 
\hline
\end{tabular}
\end{center}
\label{tb:ccb}
\end{table}%

\begin{table}
	\caption{Emission factors ($\epsilon$) and uncertainties ($\sigma$) for each sector ($s$). The emission factors (EF) are based on the GFED3 used in GEOS-Chem. The emission factors (2nd and 3rd columns) are slightly different to those reported by \cite{Werf:2010bh} (shown in brackets).  Differences are associated in brackets. Uncertainties (4th and 5th columns) are reported by \cite{Andreae:2001fk}  and \cite{Akagi:2011ve}, as standard deviations, ranges,  or mean variability. The uncertainty estimates are converted to log-normal distributions in equations 2 and 3.  CO EF are in gCOkg$^{-1}$ and CO$_2$ EF are in gCO$_2$kg$^{-1}$\label{tab:emfac}}
	\begin{tabular}{|c|c|c|c|c|c|c}\hline\hline
		Sector $s$	&  $\epsilon_{CO}$ &  $\epsilon_{CO_2}$   & $\sigma_{CO}$  & $ \sigma_{CO_2}$  & Uncertainty Source  \\ \hline
		Ag Waste EF	& 92 (94) & 1308 (1452) & 84 & 177  & \cite{Andreae:2001fk} \\ \hline
		Deforestation EF & 101 (101) & 1626 (1626) & 24 & 40 & \cite{Akagi:2011ve} \\ \hline
		Extratropical Forest EF	&106 (106)  & 1572 (1572) & 44  & 98 & \cite{Akagi:2011ve} \\ \hline
		Peat EF	& 210 (210) & 1703 (1563) & 60 & 65 & \cite{Akagi:2011ve} \\ \hline
		Savanna EF 	& 62 (61) & 1650 (1646) & 17 & 63 & \cite{Akagi:2011ve}  \\ \hline
		Woodland EF	& 82 (81) & 1638 (1703) & 32 & 71 & \cite{Akagi:2011ve} \\ \hline
	\end{tabular}
\end{table}

\begin{table}
	\caption{ Ocean carbon tendency 2011-2010 for each ocean basin. \label{tab:ocean}}
	\begin{tabular}{|ccccc|}\hline
		& Atlantic & Pacific & Indian & Southern \\ \hline
		PgC	& -0.187 & -0.008 & 0.011 & 0.017 \\ \hline
	\end{tabular} 
\end{table}
%% TWO-COLUMN TABLE

%t
%\begin{table*}[t]
%\caption{TEXT}
%\vskip4mm
%\centering
%\begin{tabular}{column = lcr}
%\tophline
%
%\middlehline
%
%\bottomhline
%\end{tabular}
%\end{table*}
%

%% The different columns must be seperated with a & command and should
%% end with \\ to identify the column brake.

%%%%%%%%%%%%%%%%%%%%%%%%%%%%%%%%%%%%%%%%%%%%%%%%%%%%%%%%%%%%%%%%%%%%%%%%%%%%%%

%% If figures and tables must be numbered 1a, 1b, etc. the following command
%% should be inserted before the begin{} command.

%\addtocounter{figure}{-1}\renewcommand{\thefigure}{\arabic{figure}a}

\end{document}